\documentclass[useAMS,usenatbib,usegraphicx]{mn2e}
\def\cf{{\rm cf.}\,}
\def\eg{{\rm e.g.,}\,}
\def\ie{{\rm i.e.}\,}
\def\etal{{\rm et al.}}
\def\etc{{\rm etc.}}

\def\p#1#2{{\partial#1\over\partial#2}}

\def\Mu{{\cal M}}
\def\Rho{{\cal P}}
\title{Two Dimensional Adiabatic Flows onto a Black Hole: I. Fluid Accretion}
\author[Roger D. Blandford \& Mitchell C. Begelman]
{Roger D. Blandford$^1$\thanks{E-mail: rdb@tapir.caltech.edu}
and Mitchell C. Begelman$^2$\thanks{E-mail: mitch@jila.colorado.edu.
Also at Department of Astrophysical and Planetary Sciences,
University of Colorado}\\
          $^1$Theoretical Astrophysics, Caltech 130-33, Pasadena, CA 
91125, USA\\
          $^2$JILA, University of Colorado, Boulder, CO 80309-0440, USA}
\begin{document}
\maketitle
\begin{abstract}
When gas accretes onto a black hole, at a rate either much less than or much
greater than the Eddington rate, it is likely to do so in an ``adiabatic''
or radiatively inefficient manner. Under fluid (as opposed to MHD) conditions,
the disk should become convective and evolve toward a state of
marginal instability. The resulting disk structure is ``gyrentropic,'' with
convection proceeding along common surfaces of constant angular momentum,
Bernoulli function and entropy, called ``gyrentropes."  We present a
family of two-dimensional, self-similar models which describes the
time-averaged disk structure.  We then suppose that there is a
self-similar, Newtonian torque and that the Prandtl number is large.
This torque drives inflow and meridional circulation and the resulting
flow is computed. Convective transport will become ineffectual
near the disk surface.
It is conjectured that this will lead to a large increase of
entropy across a ``thermal front'' which we identify as the effective
disk surface and the base of an outflow.  The conservation of mass,
momentum and energy across this thermal front permits a matching
of the disk models to self-similar outflow solutions. We then demonstrate
that self-similar disk solutions can be matched
smoothly onto relativistic flows at small radius and thin disks at
large radius. This model of adiabatic accretion
is contrasted with some alternative models that have been
discussed recently. The disk models developed in this paper should
be useful for
interpreting numerical, fluid dynamical simulations. Related principles
to those described here may govern the behaviour of
astrophysically relevant, magnetohydrodynamic disk models.
\end{abstract}
\section{Introduction}
\label{sec:int}
Recent X-ray observations of putative black
holes in the nuclei of nearby elliptical galaxies
show that they are extremely underluminous, some eight and four
orders of magnitude below the Eddington and Bondi luminosities,
respectively \citep[\eg][]{dim00,bag01,mus00}.
These observations have stimulated
a fresh look at the nature of the accretion process.  It appears
that when the hole is ``underfed,'' specifically when the mass
accretion rate is
well below the fiducial Eddington rate, $\dot m \equiv \dot M / \dot
M_E \ll 1$, where $\dot M_E=4\pi GM/c\kappa$
(with $M$ denoting the black hole mass and $\kappa$, the relevant opacity),
the radiative efficiency may be quite small, up to six orders of magnitude
smaller than the traditional value $\sim0.1 \ c^2\sim10^{20}$~erg g$^{-1}$.

One rationalisation of this observation \citep[\eg][and references therein]
{qua99a}, is that the gas accretes quite rapidly
under the action of viscous stress but the dissipated energy is taken
up almost exclusively by the ions, which do not radiate directly and which
cannot heat the electrons efficiently. In a large number of recent
publications, it has been supposed that electrons are heated minimally
in this manner and that there is a conservative inflow in which hot
ions advect essentially all of their binding energy across the black
hole event horizon.

However, these ``Advection-Dominated Accretion Flow" (ADAF) solutions
have a serious and fundamental shortcoming --- the accreting gas is
generically unbound and can escape to infinity. The reason why this
happens is that the gas is likely to be supplied with sufficient
angular momentum to orbit the hole and its inflow is controlled by
the rate at which angular momentum is transported outward.  This
angular momentum
transport, describable as a locally-acting torque, is necessarily associated
with a transport of energy.
If we attempt to conserve mass, angular momentum and energy in the flow,
we find that the Bernoulli function --- the energy that the gas
would have if it were allowed to expand adiabatically to infinity ---
is twice the local kinetic energy \citep[][ henceforth BB99]{bla99}.
In conventional accretion disks, this energy is radiated away and
the gas remains bound. However, when cooling is unimportant on the
inflow time --- we call this case ``adiabatic" by analogy with the
terminology for supernova remnants --- something else must happen to the
energy.

In an alternative description of adiabatic accretion, BB99
proposed that the inflow
is non-conservative and that the radial energy transport
drives an outflow that carries away mass, angular momentum and energy,
allowing the disk to remain bound to the hole.
(This was not a new proposal.  \citet{sha73}
were aware of this possibility and it has been discussed in
many subsequent studies.)
In these ``ADiabatic Inflow-Outflow Solutions" (ADIOS), the final
accretion rate into the hole may be only a tiny fraction (in extreme
cases $<10^{-5}$) of the mass supply at large radius (although this is
not required).  This leads to a much smaller luminosity than
would be observed from a conservative flow.
This is important from an observational perspective, because different
assumptions about the extent and nature of the outflow affect the derived
densities, temperatures, \etc, of the emitting regions, and can lead to
very different conclusions based on phenomenological fits to multi-band data.
In particular, most ADAF models posit essentially thermal emission,
whereas ADIOS models are supposed to involve nonthermal emission by
relativistic electrons as may be accelerated in the trans-sonic, shearing,
magnetized flow surrounding the black hole.
It has long been tempting \citep[\eg][]{bla78,ree82}
to associate underfed accretion onto very massive black holes with
radio galaxies and quasars, and the presence of an outflow provides a natural
agency for collimating relativistic jets, which are probably powered by
electromagnetic or hydromagnetic processes close to the event horizon.

The epitome of an underfed black hole is the Galactic Centre
\citep[\eg][]{mel01}. Here the rate of gas supply at the Bondi radius
($r\sim10^5 m$, where $m \equiv GM/c^2$) is estimated to be
$\sim10^{21}$~g s$^{-1}$ \citep[\eg][]{dim00} while the bolometric
luminosity is no more
than $\sim10^{36}$~g s$^{-1}$, giving an efficiency of conversion of mass
supply to radiant energy of $\la10^{-6}c^2$ --- hardly an advertisement
for gravity power! Subsequent observations of Sgr A$^\ast$ have shown that
the X-ray emission is rapidly variable and has a steep
spectrum \citep{bag01}, which is inconsistent with simple ADAF models
that predicted a bremsstrahlung spectrum produced far from the black hole.
Furthermore, \citet{ait00} (\cf \citet{bow02}) have measured
mm linear polarisation, which suggests that the plasma density
close to the black hole is much less than would be associated with
a conservative inflow \citep{ago00}.

As pointed out in \citet{beg82,bla85} and BB99,
the ADIOS analysis may also be appropriate for ``overfed'' accretion,
when $\dot M \gg \dot M_E$.  Here the emissivity is large enough
that radiation is emitted freely.  However, the opacity is also
large, so that the photons cannot escape on an inflow timescale
and are trapped by the flow. Hence the radiative efficiency,
defined by the ratio of the escaping luminosity to the mass
supply, is also low.  At high accretion rates, the gas
is again found to be unbound and it is proposed that inflow can
take place only in the presence of a compensating outflow. There are also
good observational reasons for believing that radiatively-driven
outflows are associated with overfed accretion.  The Galactic
source SS433 \citep{mar90,kin01} appears to be an accreting black hole from
which gas escapes at a rate at least a hundred times the critical
rate.  Galactic superluminal sources, like GRS1915+112
\citep{mir98}, also appear to be accreting rapidly and driving powerful
outflows.  Overfed, massive holes have long been associated with
radio-quiet quasars which are classified as broad-absorption line quasars
when viewed from an equatorial direction \citep[\eg][]{bla78,wey97}.
Although we do not understand enough physics to predict the maximum mass
accretion rate for an underfed disk and the minimum one for the overfed
case, the principle is clear --- the classical, thin accretion disk is only
a good description for a limited range of intermediate
mass accretion rates and may only apply to a minority of accreting
black holes.

There is now some observational evidence for the proposition that
radiation-dominated accretion flows are also ``demand-limited'' rather than
``supply-driven.'' An argument, originally due to \citet{sol82}, associates
the energy radiated by AGN (mostly quasars) with the mass of the relict
black holes.  The most recent estimate of these two quantities,
allowing for the redshift of the emitted photons and the bolometric
correction for unobserved emission \citep{yut02}, (but see \citet{fab02}),
finds that they are in the ratio $\sim 0.1-0.2c^2$. This implies that
the binding energy of the accreting gas as it crosses the event horizon
cannot be much less than $\sim 0.1$ as would be true of a radiation-dominated
ADAF.  Either black holes with $M\sim10^8{\rm M}_\odot$, which account for
most of the relict mass, acquire most of their mass during thin disk
accretion, which requires an unlikely  fine-tuning of the mass supply
rate, or most of the mass supplied is blown away. (Note
that this constraint does not require quasars to satisfy
the Eddington limit \citep{beg02}.)

BB99 presented a family of simple, one-dimensional similarity solutions
that span a large range of allowed flows, parametrised by the
rates at which mass, angular momentum and energy are extracted. There
was no discussion of the extraneous physical considerations that would allow
one to determine these parameters within a broad range limited only
by general thermodynamic and mechanical considerations.
In this paper, we present a more detailed, fluid dynamical description
of ADIOS disks that exhibits the manner by which the global transport of
mass, angular momentum and energy might depend upon the microphysics
assumed. We
explicitly ignore magnetic field in this paper, so the solutions presented
below are not directly applicable to observed disks but they are  useful
for bringing out salient principles. They may also aid in
analyzing numerical simulations.

In particular, we base our models on the prediction
that adiabatic fluid disks are convective \citep{bar73,pac82,beg82,bla85}.
In more recent developments, \citet{qua99a}
and \citet{nar00} have proposed that convection transports
energy outward while carrying angular momentum inward \citep[\cf][]
{ryu92,bal00}. This type of flow has been styled a Convection-Dominated
Accretion Flow or ``CDAF.'' In analytic models of a CDAF, the inward radial
transport of angular momentum by convective motions
exactly cancels the outward
transport by viscous torque and the net mass accretion rate is very small.

In this paper, we adopt a quite different model for the convection.
Following \citet{bar73,pac82,beg82} and \citet{bla85}, we argue
that the
convective transport is vertical rather than radial, consistent
with the H\o iland criterion.  When convection is efficient, it implies
that the disk structure is ``gyrentropic'' --- that is to say the isentropes
coincide with ``isogyres'' (surfaces of constant specific angular momentum).
This is the state of marginal, convective instability and it
leads to the transport of mass, angular momentum and energy
to the disk surface where these quantities can be removed by the outflow.
This prescription allows us to compute two-dimensional, hydrostatic
disk models. However, these models do not describe inflow. We therefore add
an explicit, though small, viscosity that leads to a torque
across the gyrentropes. We show that this torque must also drive
a meridional circulation which can be computed. These circulating
disk models are accurate only in the limit of small viscosity, when
the convection can be efficient almost to the disk surface, just as happens
in solar-type stars. The viscosity in accretion disks is not now thought
to be so small and, as a consequence, the outflow can have a significant
impact on the disk structure. We therefore further modify our circulation
disk models by truncating them at a ``thermal front'' where the convective
energy flux is transformed into heat so that the associated pressure can
self-consistently drive an outflow to infinity.  In the case of an
ion-dominated flow, the region of the disk
downstream of the thermal front may be identified with an active corona.

One important difference between adiabatic accretion disks and their
conventional, radiative counterparts is that, as they are thick, with
opening angles $\sim1$, the distinction between the thermal timescale
and the viscous timescale \citep[\eg][]{fra01} is lost.  The outflow
adjusts on a timescale $\sim O(\Omega^{-1})$, where $\Omega$
is the angular frequency, whereas, as long as it is dynamical
stable, the disk structure changes on a longer timescale, $\sim O(\alpha^{-1}
\Omega^{-1})$, where $\alpha$ is the conventional viscosity parameter.
These considerations will prove to be important when we
discuss how a disk relaxes to a particular configuration in response
to its assumed microphysical properties.

In the following section,
we generalize the one-dimensional treatment of disk accretion in
BB99 to accommodate alternative equations of state and introduce six models
that span the types of flows that can be described by our solutions.
In \S~\ref{sec:two}, we discuss two-dimensional convective stability
in a rotating disk and derive models of two-dimensional gyrentropic
disks in hydrostatic equilibrium.
We next introduce a Newtonian, viscous stress that
drives inflow and meridional circulation and supplies mass,
angular momentum and energy to the disk surface (\S~\ref{sec:cir}).
Finally, these disk models are modified to match self-similar
outflows (\S~\ref{sec:out}).
A legitimate concern about self-similar disk models is that
they may be invalid because they must
fail at large and small radii. In \S~\ref{sec:nss}, we
demonstrate that our self-similar solutions can be matched onto a
general relativistic flow close to the hole, and to a thin disk
near an outer, transition radius. This is followed in 
\S~\ref{sec:alt} by a critical comparison with
some alternative descriptions of adiabatic accretion that have appeared
in the recent literature. We summarize our main conclusions in
\S~\ref{sec:dis}. We shall discuss the more relevant
problem of magnetic accretion in Paper II and the application to
selected astronomical sources in Paper III.

\section{One-dimensional Disks}
\label{sec:one}

\subsection{Conservation Laws}
\label{ssec:cons}

In BB99, we gave a simple explanation of why conservative,
adiabatic accretion disk flows are
unbound. Specifically, we showed, using a one-dimensional model,
that the Bernoulli function, defined by
\begin{equation}
\label{berndef}
B=H+{\Omega^2R^2\over2}-{1\over R},
\end{equation}
equals $\Omega^2R^2$ when the total energy and angular momentum
fluxes vanish. Here, $\Omega$ is the angular frequency,
$H=\gamma P/(\gamma-1)\Rho$ is the enthalpy per unit mass, $P$ is the
pressure, $\Rho$ is the density
and $GM$ has been set to unity. Radiation-dominated and gas-dominated
accretion correspond to adiabatic exponents $\gamma=4/3,5/3$,
respectively. A positive Bernoulli function implies that an
element of gas already has enough internal energy, after expanding
adiabatically and doing work on its surroundings, to escape to
infinity.  Before we discuss possible mechanisms for effecting this
removal, we must reprise and generalise the 1D results.

For convenience, we introduce an entropy function
\begin{equation}
\label{entdef}
S=P^{1/\gamma}/\Rho,
\end{equation}
which is monotonically related to the true thermodynamic entropy
in thermal equilibrium. (We shall not require that the gas be
in local thermodynamic
equilibrium, only that $P\propto\Rho^\gamma$ when an element of gas changes
its density in such a manner that there is negligible
dissipation and heat exchange with its surroundings.)

For the moment we ignore
vertical gradients, and suppose that the flow is
stationary. (The assumption of stationarity need not seriously restrict our
conclusions, if they are applied to the time-averaged flow.) The disk
is hypothesised to evolve under a combination of internal torque
and external loss of mass, angular momentum and energy.
This allows the remaining gas
to flow radially inward at a rate small compared with the rotational
speed. In the absence of a better prescription and in order to
elucidate general principles, we adopt, initially, a self-similar disk
mass {\em inflow}
\begin{equation}
\label{wdotm}
\dot M=-\Mu V_r\propto R^n;\qquad0\le n<1,
\end{equation}
where $\Mu$ is the mass per unit radius and $\Mu,V_r,n$ substitute
for $\mu,-v,p$ of BB99,
respectively. The reason for the restriction $n\ge0$ is that gas is
only supposed to leave the disk and for the inequality $n<1$ is that
the energy flowing outward through the disk
presumably decreases with radius as it is released mostly at small
radius and is carried off by an outflow. We can also use
eq.~(\ref{wdotm}) to define the mass {\it loss} per unit radius
\begin{equation}
\label{dwdotm}
J\equiv{d\dot M\over dr}=n{\dot M\over r}.
\end{equation}
Here, and in what follows, we shall treat the disk and outflow as
symmetric about the equatorial plane, so that all integral quantities
refer to one hemisphere.

We next suppose that the disk angular momentum {\em inflow} also
varies as a power law
\begin{equation}
\label{fldef}
F_L=\dot M(L-{\cal G})=\lambda\dot MR^{1/2}={2n(1+\eta)\dot ML\over1+2n}
\end{equation}
where $L=R^2\Omega$ is the specific angular momentum of the disk
and ${\cal G}$ ($\equiv G$ in BB99) is the
generalised, internal torque per unit $\dot M$ and is assumed to be
positive. As we discuss further below, this
equation can be regarded
as a definition of the torque but it has to be interpreted carefully
in the presence of convection.  We replace the parameter $\lambda$
of BB99 with a new parameter $\eta$. With this definition, the specific
angular momentum of the gas removed is
\begin{equation}
\label{dfldef}
{dF_L\over d\dot M}=(1+\eta)L.
\end{equation}
Gas dynamical outflows that exert no reaction torque on the disk have
$\eta=0$.  However, if there are magnetic fields present, as we shall
discuss explicitly in Paper II, then $\eta>0$ or if the outflow
originates below the disk surface, as we discuss in \S5, then
$\eta<0$ and so we shall retain this
generality. Provided that this torque
can be regarded as a local variable, then the second law of thermodynamics
requires that it oppose the velocity shear \citep[\eg][]{lan59}
so that
\begin{equation}
\label{etalim}
0\le\eta<{1\over2n}.
\end{equation}
However, angular momentum transport need neither be local in this
sense \citep[\eg][]{ryu92,bal00,nar00,qua00a}
nor necessarily describable in the language of
fluid mechanics \citep{qua00b}).
Furthermore, the first inequality in eq.~(\ref{etalim})
is not strictly required when there
is internal circulation, as we shall discuss below.

In a similar fashion, we assume a self-similar variation of the disk energy
{\em outflow}  and replace the energy parameter
$\epsilon$ of BB99 with the dimensionless parameter
$\beta$ according to
\begin{equation}
\label{fedef}
F_E=\dot M({\cal G}\Omega-B)={\epsilon\dot M\over R}=
{-n(\beta-1)\dot MB\over(1-n)}.
\end{equation}
The quantity $\dot M{\cal G}\Omega$ represents the mechanical work
performed by the
generalized torque from eq.~(\ref{fldef}), while $\dot M B$ is the energy
advected inward by the gas. With this definition of $\beta$, we have that
\begin{equation}
\label{dfedef}
{dF_E\over d\dot M}=(\beta-1)B,
\end{equation}
in parallel to eq.~(\ref{dfldef}).
There may be additional contributions to the
energy flux, particularly associated with convection, hydromagnetic
wave transport and thermal conduction. We shall include the first of these
below.

We can combine equations (\ref{dfldef}) and (\ref{dfedef}) to obtain
the useful relations
\begin{equation}
\label{dmdef}
{\cal G}=\left({1-n\beta\over1-n}\right){BR^2\over L}
=\left({1-2n\eta\over1+2n}\right)L.
\end{equation}
Bound disks with $B<0,{\cal G}>0$ require a minimum energy outflow with
\begin{equation}
\label{betalim}
\beta>{1\over n}>1.
\end{equation}

In the limit of a thin disk, $B\rightarrow-1/2R, L\rightarrow R^{1/2}$ and
\begin{equation}
\label{betathin}
\beta\rightarrow{3-4n(1-n)\eta\over n(1+2n)}.
\end{equation}
This additional, lower limit on $\beta$ imposes an additional,
lower limit on $\eta$
\begin{equation}
\label{etamin}
\eta>{3-n(1+2n)\beta\over4n(1-n)}.
\end{equation}
In simple, fluid models with $\eta=0$,
\begin{equation}
\label{betamin}
\beta\ge\beta_{{\rm min}}\equiv{3\over n(1+2n)}.
\end{equation}

As explained in BB99, the self-similar scalings for
pressure and density are $P\propto R^{n-5/2},\Rho\propto R^{n-3/2}$,
which transform the approximate radial equation of motion into the form
\begin{equation}
\label{omr}
\Omega^2R^2-{1\over R}+(5/2-n){P\over\Rho}=0.
\end{equation}
Combining with eq.~(\ref{berndef}), we obtain
\begin{equation}
\label{lsolve}
L=\left[{2(5-2n)[1 - (\gamma-1)BR]-2\gamma(3-2n)]\over5-2n+\gamma(2n-1)}
\right]^{1/2}R^{1/2} .
\end{equation}
Combining eq.~(\ref{lsolve}) with eq.~(\ref{dmdef}) allows us to solve for
$L,B,{\cal G}$ as explicit functions of $n,\beta,\eta$ subject
to inequalities (\ref{wdotm}), (\ref{etalim}), (\ref{betalim}), and
(\ref{etamin}).

However, in this paper we shall follow a different approach. Instead of
emphasizing the full range of outflow models that might be possible,
we shall explore the manner in which the local physics might dictate
a particular, global solution. It is then more convenient to express
$n,\beta,\eta$ in terms of the fluid variables that describe the disk.
A convenient, general choice is $L,B,{\cal G}$ and we find that
\begin{equation}
\label{nlb}
n(L,B)={(5-\gamma)L^2/R+10(\gamma-1)BR+2(3\gamma-5)
\over2(\gamma-1)(2+2BR-L^2/R)}
\end{equation}
independent of ${\cal G}$. Lengthier expressions can be derived
for $\beta(L,B,{\cal G})$ and  $\eta(L,B,{\cal G})$,
which will not be reproduced. For the special case of interest
here, $\eta=0$, it is simplest to solve eq.~(\ref{dmdef})
for $\beta$
\begin{equation}
\label{betlb}
\beta(L,B)={(1+2n)B R-(1-n)L^2/R\over n(1+2n)BR},
\end{equation}
and substitute eq.~(\ref{nlb}). In this case, $\beta$ is also independent of
${\cal G}$.
\subsection{Illustrative Models}
\label{ssec:1dmod}
\begin{figure*}
\includegraphics[width=18cm]{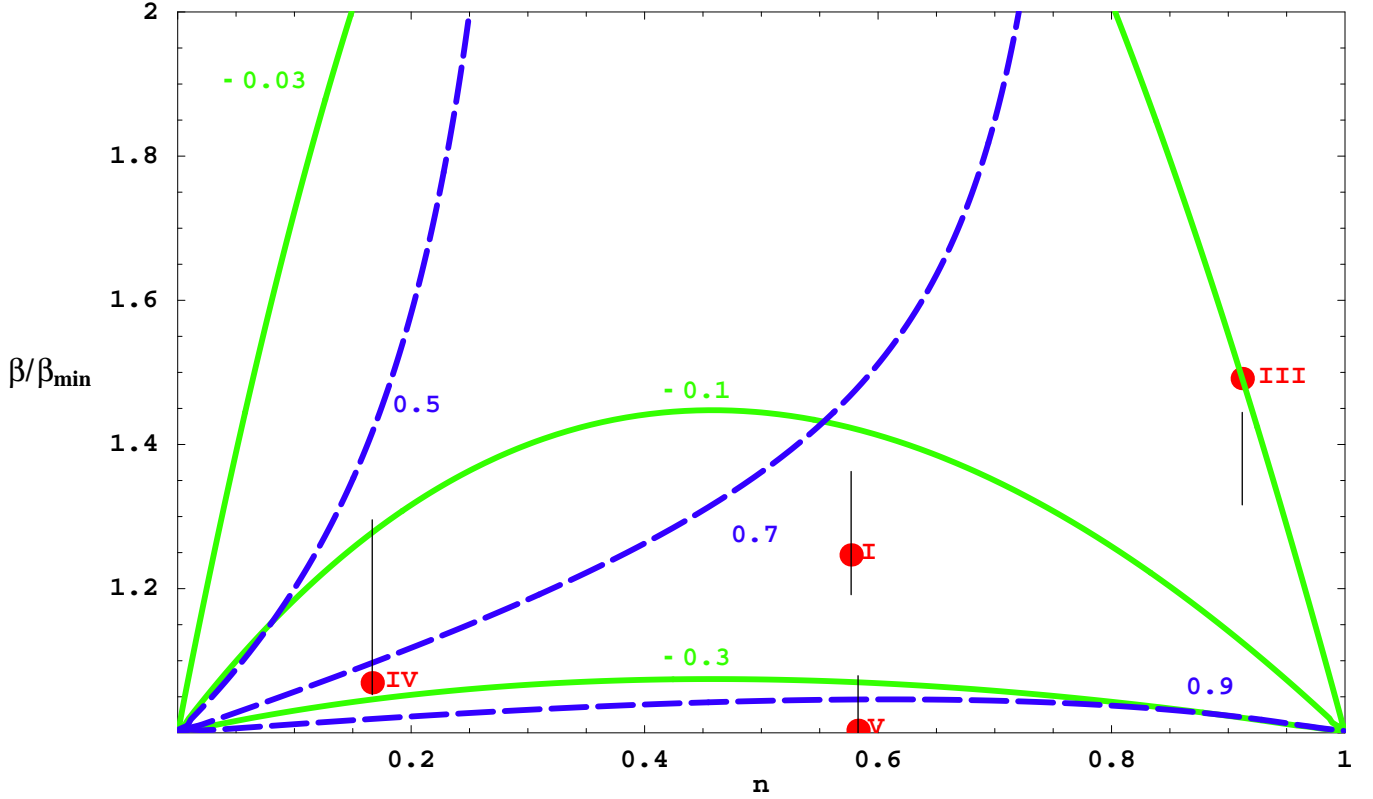}
\caption{Four disk models of self-similar, gyrentropic accretion
discussed in the text and located on the $n-\beta/\beta_{{\rm min}}$
plane assuming a fluid disk with $\eta=0, \gamma=5/3$. (Model II has
$\gamma=4/3$; Model VI has large $\beta$.)
The ordinate is the ratio of the scaled energy outflow to that
required for a thin disk with $L=R^{1/2},B=-0.5/R$.
The dashed lines are contours of constant $L/R^{1/2}$ in the case of 1D models
and $\ell_0$ for 2D models; the solid lines are contours of constant
$BR$ for 1D models and $b_0$ for 2D. The points correspond to the values of
$\beta/\beta_{{\rm min}}$ for 1D models.  The vertical
lines passing upward through these points connect the equivalent values
of $\beta/\beta_{{\rm min}}$ for circulation models to those for the
outflow models (\cf Table~\ref{tbl:mod}).
Using these and other measures, it can be shown that the 1D models
provide a surprisingly good representation
of the 2D models.}
\label{fig:nbe}
\end{figure*}

In this paper we shall use six examples of combined disk-wind flows
to allow us to
explore a range of models described by our approach and which illustrate some
more general principles. We introduce here the
one-dimensional versions of these models which we shall shortly describe
in two dimensions. The principal 1D and 2D characteristics of these models
are listed in Table~\ref{tbl:mod} and plotted for four of these
models, with $\eta=0$ and $\gamma=5/3$, in Fig.~\ref{fig:nbe}.
\begin{table*}
\begin{tabular}{lcccccccccccccccccccc}
\small
All&&&&\vline&1D&&\vline&2D&\vline&Circ.&\vline&Out.&&&&\vline&Wind\\
\hline
No.&$\gamma$&$\ell_0$&$b_0$&
\vline&$n$&$\beta$&
\vline&$\theta_d$&
\vline&$\beta$&
\vline&$\beta$&$\eta$&$b_d$&$\theta_w$&
\vline&$b_w$&$v_\infty$&$\theta_j$\\
\hline
I&1.67&0.75&-0.15&\vline&0.58&3.01&\vline&0.42&\vline&2.86&
\vline&3.27&-0.10&-0.31&0.64&\vline&0.62&0.77&0.39\\
II&1.33&0.90&-0.20&\vline&0.58&3.12&\vline&0.60&\vline&2.87&
\vline&3.26&-0.09&-0.30&0.91&\vline&0.59&0.89&0.39\\
III&1.67&0.75&-0.03&\vline&0.91&1.74&\vline&0.18&\vline&1.54&
\vline&1.69&-0.03&-0.19&0.36&\vline&0.11&0.19&0.32\\
IV&1.67&0.75&-0.25&\vline&0.17&14.4&\vline&0.56&\vline&13.9&
\vline&17.6&-0.76&-0.37&0.77&\vline&5.60&2.74&0.66\\
V&1.67&0.99&-0.46&\vline&0.58&2.38&\vline&1.34&\vline&2.33&
\vline&2.59&-0.14&-0.49&1.40&\vline&0.55&1.13&1.34\\
VI&1.67&0.63&-0.02&\vline&0.58&10.0&\vline&0.11&\vline&7.05&
\vline&8.25&-0.09&-0.11&0.34&\vline&0.78&0.47&0.24\\
\end{tabular}
\caption{Physical parameters that characterise the six disk-outflow models,
identified by Roman numerals. The models are described
in four formalisms of increasing sophistication: a 1D model,
a 2D gyrentropic model extending to zero pressure and density,
the same with the internal circulation and convective heat flux
computed self-consistently (Circ.), and a disk whose surface is chosen
to be a thermal front that coincides with the base of the outflow (Out).
The models are parametrised by the specific heat ratio $\gamma$,
the equatorial Bernoulli function $b_0$ ($=RB$ for 1D models)
and the scaled, equatorial angular momentum
$\ell_0$ ($=R^{-1/2}L$ for 1D models).
These parameters are sufficient to derive the mass loss exponent $n$
and the energy loss exponent $\beta$ in 1D, assuming that $\eta=0$.
The basic 2D treatment adds a disk opening angle
$\theta_d$. A viscous torque is introduced in the circulation
models which modifies the derived energy loss rate.
Further modifications to $\theta_d,\beta,\eta$ occur in the outflow models.
The scaled Bernoulli function changes from $b_d$ to $b_w$ on crossing
the thermal front (at an angle $\theta_w$)
in the outflow models. Asymptotically, the wind
flows on the surface of an evacuated cone with scaled
speed $v_\infty$ and opening angle $\theta_j$.}
\label{tbl:mod}
\end{table*}
\subsubsection{Model I: Fiducial, ion-supported disk}
\label{sssec:fid}
This reference model assumes that the hole is underfed so that the
gas pressure is ion-dominated (with
$\gamma=5/3$). We suppose that the gas has moderate pressure (parametrised
by $B=-0.15/R$) and centrifugal support
(parametrised by $L=0.75R^{1/2}$) and consequently moderate mass and energy
loss rates ($n\sim0.6,\beta\sim3$).
\subsubsection{Model II: Radiation-supported disk}
\label{sssec:radn}
Overfed, radiation-dominated accretion with $\gamma=4/3$.
$L,B$ are chosen to give similar values of $n,\beta$ to Model I.
\subsubsection{Model III: Thick disk}
\label{sssec:thick}
A much less bound version of Model I ($B\sim-0.03/R$)
but with a similar circular velocity. The 2D disk is very thick with a
small opening angle, $\theta\sim0.2$. The mass loss rate is high ($n\sim0.9$)
and so the energy loss parameter is low ($\beta\sim1.7$).
\subsubsection{Model IV: Intermediate disk}
\label{sssec:thin}
A more tightly bound version of Model I ($B=-0.25/R$), with a
somewhat thinner disk.
The mass loss rate is relatively small ($n\sim0.2$) and so $\beta\sim14$
has to be correspondingly large to compensate.
\subsubsection{Model V: Fast disk}
\label{sssec:slow}
A fast, thin disk ($L\sim0.99R^{1/2}$). In order to create a model to
explore this
limit we choose to keep the mass loss rate index $n$ equal to that
of Model I.
\subsubsection{Model VI: Slow disk}
\label{sssec:lowdotm}
A slowly rotating, thick disk ($L\sim0.6R^{1/2}$), that is very weakly bound
($B\sim-0.02/R$), but with similar mass loss index
to that of Model I.
Again $\beta\sim10$ is large.

\section{Two-dimensional Disks}
\label{sec:two}

\subsection{H\o iland Criteria}
\label{ssec:gyrent}

In order to make a two-dimensional model of a slowly accreting and
consequently hydrostatic disk, we must specify some
relationship among the thermodynamical variables $P,\Rho,S,$
etc. Our choice depends upon considerations of
stability. An adiabatic, fluid accretion disk naturally develops a negative,
radial entropy gradient as heat is generated in its interior. If rotation
were unimportant, it would become unstable according to the
Schwarzschild criterion.  However, a Keplerian disk has a positive
angular momentum gradient and, if we were to ignore entropy, it would be
stable according to the Rayleigh criterion. For thin disks, the rotational
stabilisation dominates the entropy destabilisation.  However, for
thick disks, the two effects must be compared directly.

To do this, consider a small, flat ribbon of fluid with an azimuthal length
much greater than its width
which, in turn, is much greater than its
thickness. Let the ribbon undergo a displacement
in the poloidal plane, parallel to its width. Assume that the motion is
sufficiently slow that the ribbon remains in pressure equilibrium
with its surroundings and sufficiently rapid that its entropy
$S$ is unchanged \citep[\eg][]{gol67,tas78,beg82}. The net buoyant
acceleration on the ribbon is then given by
\begin{eqnarray}
\label{accb}
\delta\vec a_{{\rm buoy}}&=&-(\delta\vec r\cdot\nabla)S \ \left({\p{\ln\Rho}S}
\right)_P\left({\nabla P\over\Rho}\right)\nonumber\\
&=&(\delta\vec r\cdot\nabla)S \ \nabla\left({P^{1-1/\gamma}
\over1-1/\gamma}\right),
\end{eqnarray}
where the spatial gradients are in the surrounding medium.
If the displacement is also rapid enough for the ribbon's
specific angular momentum $L$ to be unchanged during the displacement,
the surplus centrifugal acceleration of the ribbon relative to its
surroundings is likewise given by
\begin{equation}
\label{accc}
\delta\vec a_{{\rm cent}}=-(\delta\vec r\cdot\nabla)L^2\left({\vec R
\over R^4}\right)=(\delta\vec r\cdot\nabla)L^2\nabla\left({1\over2R^2}\right).
\end{equation}
The total acceleration is the sum of equations (\ref{accb}) and (\ref{accc}).

If a virtual displacement $\delta\vec r$ is made, the virtual work
done, per unit mass of fluid, is $\delta W=U_{ij}\delta r_i\delta r_j/2$,
where the tensor ${\rm U}$ is given by
\begin{equation}
\label{virwk}
{\rm U}=\nabla\left({P^{1-1/\gamma}\over1-1/\gamma}\right)
\otimes\nabla S+\nabla\left({1\over2R^2}\right)\otimes\nabla L^2.
\end{equation}
Only the symmetric part of ${\rm U}$ need be retained and we can rotate
axes in the $r-\theta$ plane so that it is diagonal.
The flow will be unstable if there exists a displacement $ \delta\vec
r$ such that
$\delta W>0$. If the trace of ${\rm U}$ is positive, then there must be
unstable displacements. This leads to the first H\o iland instability
condition,
\begin{equation}
\label{hoione}
\nabla\left({P^{1-1/\gamma}\over1-1/\gamma}\right)\cdot\nabla S
+\nabla\left({1\over2R^2}\right)\cdot\nabla L^2>0.
\end{equation}
Disks that satisfy this inequality are unstable in the equatorial plane
to radial displacements. If we adopt our self-similar scalings and
impose hydrostatic balance in the equatorial plane, then this
instability condition can be re-written as
\begin{equation}
\label{adef}
R^3\Omega^2<{3\gamma-5-2n\gamma+2n\over\gamma-5-2n\gamma+2n}\equiv a(\gamma,n).
\end{equation}
For $\gamma=5/3\ (4/3)$, $a(\gamma,n)$ increases from $0\ (3/11)$ to
$2/7\ (5/13)$ as $n$ increases from 0 to 1. It turns out that all of the
1- and 2-dimensional models discussed below strongly violate
inequality (\ref{adef}). For this reason, we argue that adiabatic,
fluid disks are {\it quite stable to radial convection}.

Nonetheless, when ${\rm Tr(U)}<0$, a second H\o iland criterion
must be considered and there will still be a range of unstable
displacement directions if
\begin{equation}
\label{hoitwop}
\left[\nabla\left({P^{1-1/\gamma}\over1-1/\gamma}\right)\times\nabla\left(
{1\over2R^2}\right)\right]\cdot\left[\nabla S\times\nabla L^2\right]<0
\end{equation}
or
\begin{equation}
\label{hoitwo}
(\nabla P\times\nabla R)\cdot(\nabla S\times\nabla L)>0.
\end{equation}
In our application, the onset of instability is always associated with a
change in sign of $\nabla S\times\nabla L$, not $\nabla P\times\nabla R$.

\begin{figure}
\includegraphics[width=8cm]{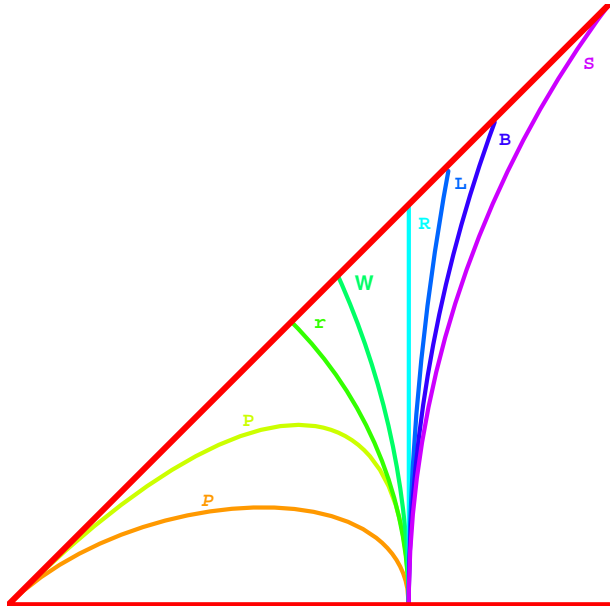}
\caption{Structure of convectively-stable accretion disks.
Level surfaces of density ($\Rho$), pressure ($P$),
spherical radius ($r$), angular frequency ($\Omega$),
cylindrical radius ($R$),
angular momentum ($L$), Bernoulli function ($B$) and entropy ($S$)
passing through an equatorial ring of radius $r$. The configuration
shown is dynamically stable according to both H\o iland
criteria eq.~(\ref{hoitwo}). When the
order of the isentropes and isogyres is reversed, the flow becomes
convectively unstable.}
\label{fig:hoi}
\end{figure}

In order to interpret inequality (\ref{hoitwo}), it is useful to define
a series of 2D
surfaces that are tangent on an equatorial ring lying in the disk
midplane and on which $S, B, L, R, \Omega, r, P, \Rho$ are constant
(Fig.~\ref{fig:hoi}).
It is straightforward to show, by combining the
H\o iland criteria with the equation of hydrostatic equilibrium and
its curl, that the surfaces of constant $S,L,B,R,\Omega,r,P,\Rho$,
must be nested in order of increasing Gaussian curvature in a stable,
fluid disk \citep{beg82, bla85}. (Actually, it is
not formally required that the ``isorotes'' [$\Omega=$ constant surfaces]
be less curved than spheres, $r=$ constant, though this is true for
all of our solutions.)

The surfaces on which the Bernoulli function $B$ is constant
are particularly important. We call these ``isoberns.''  Using the
equation of hydrostatic equilibrium, it can be shown that
\begin{equation}
\label{gradbern}
\nabla B=H\nabla\ln S+\Omega\nabla L.
\end{equation}
Equation (\ref{gradbern}) implies that the isoberns
lie between the ``isogyres'' (surfaces of constant $L$) and the
isentropes. Stability
requires that the isogyres lie inside the isentropes (eq.~[\ref{hoitwo}]).
When the disk is marginally stable, $S,L,B$
are constant on a common surface, which we call a ``gyrentrope.''
We argue that adiabatic, fluid disks should evolve quickly towards this state
and we call such disks ``gyrentropic'' \citep[\cf][]{bar73}.
In a marginally unstable disk, the growing eigenmodes have
displacements that lie between the isentropes and the
isogyres. In other words, near marginal stability, thin ribbons of gas
move in opposite directions roughly along the nearly coincident
isentrope/isogyre/isobern $=$ gyrentropic surfaces.

\subsection{Numerical Simulations}
\label{ssec:numsim}

There have been many simulations of fluid dynamical accretion disks,
both two- and three-dimensional. In particular, \citet{sto99}
(SPB99) have carried out
two-dimensional simulations of adiabatic disks evolving under a
variety of prescriptions.  Gas was supplied at an intermediate radius and
endowed with viscosity.
It spread inward (and outward) on a viscous timescale and
became strongly convective, developing a gyrentropic structure at
intermediate radii. Convection transported energy and
angular momentum primarily along gyrentropes.  In directions normal to
the gyrentropes, the only means of transporting energy and angular momentum
were via the viscous stress and advection.  Since viscosity transports
angular momentum outward with a positive divergence in the equatorial
regions, the loss of angular momentum was balanced by a combination
of meridional flow and convective transport of angular momentum between
high latitudes and the equatorial region.
When the gas was marginally unstable according to criterion
eq.~(\ref{hoitwo}), convection exchanged mass,
energy and angular momentum between the disk interior and high latitude
and not radially outward throughout the disk.

\begin{figure}
\includegraphics[width=8cm]{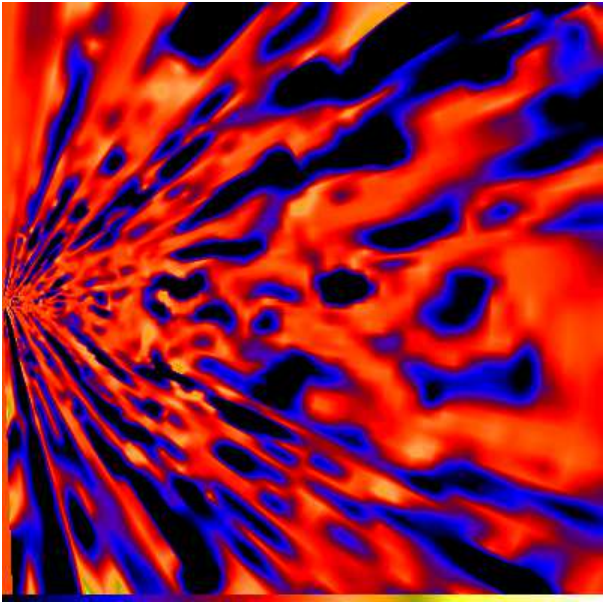}
\caption{Cross-correlation of specific angular momentum and specific
entropy from Run K of SPB99, showing convection along gyrentropic
surfaces. A time-sequence of frames from this simulation shows both
inflow and outflow of fluid elements parallel to these surfaces,
particularly apparent at mid- and high latitudes.}
\label{fig:spb}
\end{figure}

The details of this balance between convection
and circulation in the simulations depended upon the form of the
viscous stress.
For example, when the viscosity $\nu$ was set to be proportional to
$\rho$ (Run B of SPB99), the viscous torque was balanced partially by
inflow in the equatorial zones.  But at high latitudes, where the
gyrentropic surfaces themselves became roughly radial, the angular
momentum flux was dominated by convective transport and the mass flux
was generally outward, producing a quadrupolar flow pattern. When the
viscosity was chosen to satisfy the self-similar scaling law,
$\nu\propto r^{1/2}$, (Run K of SPB99), convection along gyrentropes
was strikingly demonstrated in the cross-correlation of the entropy
and the angular momentum (Fig.~\ref{fig:spb}).

SPB99 measured separately the radial mass inflow and outflow rates and
found that they both obey $\dot M\propto r$ very roughly. The net mass
accretion rate, given by the inflow rate minus the outflow rate, is
approximately independent of $r$ and equals
the inflow at the inner radius. In other words, the mass reaching the hole
at $r_{{\rm in}}$ is a fraction $\sim(r_{{\rm in}}/r_{{\rm out}})$
of that supplied at $r_{{\rm out}}$. This is a strong indication that the
flows had not become stationary which would probably require
a continuous supply of gas at large radius.

\subsection{Global Consequences of the Gyrentropic Hypothesis}
\label{ssec:global}

We have argued, on the basis of the
one-dimensional treatment, that a rotating flow will have $B>0$ unless
there is some means of removing energy. We have also argued that
such a flow will become convective and that the natural direction of
energy transport is along gyrentropes, where $B$ is constant, to the
disk surface.
This strongly suggests that adiabatic, fluid disks lose energy though
some form of outflow from their surfaces.

However, as emphasized by \citet{abr00}, the condition that
$B>0$ somewhere does not automatically imply outflow. Consider a general,
axisymmetric disk with surface described by the equation $r=r_d(\theta)$.
As with a star, the surface properties of a disk
can be quite subtle, but it is
reasonable to suppose that the enthalpy at the top of the convection zone
is much smaller than in the interior and can be ignored.
In addition, the pressure will be small; consequently, the surface will
be an isobar. Resolving the equation of motion along the disk surface,
we obtain a differential equation that relates its shape to the variation
of $B$ on the surface (and, indirectly, within the disk):
\begin{equation}
\label{disksurf}
{d\ln r_d\over d\ln\sin\theta}={-2(1+Br)\over1+2Br} .
\end{equation}
 From the form of this equation, we discover that the disk surface can
have $B>0$ provided that $d\ln\sin\theta/d\ln r_d<-1/2$, that is to say
within a funnel where the cross section increases with height slower
than parabolically. However, outside such a
funnel, the surface must have $B<0$ if
it is to remain bound. Indeed, when the disk latitude decreases with $r$,
as it must eventually, then $-1<Br<-1/2$ must be satisfied along the
free surface. Note, especially, that if
the disk surface is conical, as it is in our similarity solutions, then
\begin{equation}
\label{conebern}
B(r,\theta_d)={-1\over2r}
\end{equation}
over the entire surface.

The requirement that $B<0$ on the surface of a stable
disk (excluding a funnel),
however, is inconsistent with having $B>0$ in the interior. This is because
the second H\o iland criterion, eq.~(\ref{hoitwo}),
requires that $B$ increase with altitude.
It is for these reasons that we argue that conservative, adiabatic, fluid
accretion disks do not exist and, instead, there must always be outflows.
\subsection{Self-Similar, Gyrentropic Disks}
\label{ssec:gyr}
We now present an elementary description of two-dimensional, self-similar,
adiabatic, gyrentropic fluid disks. We shall make two key approximations
in this description --- that the only motion is rotational and that the disks
are gyrentropic all the way to their surfaces where the density and pressure
vanish simultaneously. Both of these approximations become
accurate in the limit that the rate of dissipation tends to zero.
In \S~\ref{sec:cir}, we shall introduce a prescription
for relaxing the first of these and in \S~\ref{sec:out}, we shall address
the second.

In two dimensions, our assumption of self-similarity requires that the
pressure, density and specific angular momentum satisfy the scalings:
\begin{equation}
\label{pscale}
P=r^{n -5/2}p(\theta),\ \ \Rho=r^{n - 3/2}\rho(\theta),
\ \ L=r^{1/2}\ell(\theta) .
\end{equation}
(Note that we consistently use upper case for the physical variables and
lower case for their angular variations.)

The equations of motion are
\begin{eqnarray}
\label{euler}
{1\over\Rho}{\partial P\over\partial r}&=&-{1\over r^2}+
{L^2\csc^2\theta\over r^3}\\
\label{eulert}
{1\over\Rho r}{\partial P\over\partial\theta}&=&
{L^2\cot\theta\csc^2\theta\over r^3},
\end{eqnarray}
from which we obtain
\begin{eqnarray}
\label{eulersim}
(5/2-n)p/\rho&=&1-\ell^2\csc^2\theta\\
\label{eulersimt}
p'/\rho&=&\ell^2\cot\theta\csc^2\theta,
\end{eqnarray}
where the prime denotes differentiation with respect to $\theta$.

Now consider the entropy function
\begin{equation}
\label{entropy}
S=P^{1/\gamma}\Rho^{-1}=s(\theta)r^{{-a\over1-a}},
\end{equation}
where $a(\gamma,n)$ is defined in eq.~(\ref{adef}) and
$s=p^{1/\gamma}\rho^{-1}$.

Gyrentropicity implies that $S=S(L)$ and so self-similarity requires that
\begin{equation}
\label{entscale}
S\propto(r^{1/2}\ell)^{{-2a\over1-a}}.
\end{equation}
This provides an algebraic relation among $p$,$\rho$, and $\ell$.

The solution to equations (\ref{eulersim}), (\ref{eulersimt}) and
(\ref{entscale}) is
\begin{eqnarray}
\label{psol}
\label{ellsol}
\ell&=&\left\{a+[a^2+\ell_0^2(\ell_0^2-2a)
\csc^2\theta]^{1/2}\right\}^{1/2}\sin\theta\\
p&=&\left({1-\ell^2\csc^2\theta\over5/2-n}\right)^{\gamma\over\gamma-1}
s^{-\gamma\over\gamma-1}\\
\label{rhosol}
\rho&=&\left({1-\ell^2\csc^2\theta\over5/2-n}\right)^{1\over\gamma-1}
s^{-\gamma\over\gamma-1}\\
\label{ssol}
s&=&s_0\left({\ell\over\ell_0}\right)^{-2a\over1-a}
\end{eqnarray}
where $\ell_0=\ell(\pi/2)$, \etc, measures the angular momentum,
\etc, in the midplane at radius $r_0$.
Equivalently, we can write
\begin{equation}
\label{lsdef}
S=s_0\left({L\over\ell_0}\right)^{{-2a\over1-a}}.
\end{equation}
The disk terminates at a free surface where $p,\rho\rightarrow0$
and so the disk opening angle is given by
\begin{equation}
\label{ellsurf}
\sin\theta_d=l_d=\ell_0\left({\ell_0^2-2a\over1-2a}\right)^{1/2} .
\end{equation}

We next use eq.~(\ref{ellsol}) to solve for the gyrentropes, $L=$~const.
These are given by
\begin{equation}
\label{gyren}
r=r_L(\theta)=r_0\ell_0^2/\ell^2.
\end{equation}
The gyrentropes intersect the disk surface at a radius
\begin{equation}
\label{disksur}
r=r_d=r_0(1-2a)/(\ell_0^2-2a).
\end{equation}

The Bernoulli function, which is also constant on gyrentropes,
can be calculated using eq.~(\ref{berndef}) as
\begin{equation}
\label{bernsol}
b=rB=-{\ell_d^2\over2\ell^2},
\end{equation}
consistent with eq.~(\ref{lsdef}).
Eq.~(\ref{bernsol}) also allows us to relate the value of the
Bernoulli function at the midplane ($\theta=\pi/2$) to the midplane
angular momentum:
\begin{equation}
\label{bmidp}
b_0=-{\ell_0^2-2a\over2(1-2a)}.
\end{equation}
Equivalent to eq.~(\ref{lsdef}), we have
\begin{equation}
\label{bldef}
B=b_0\left({L\over\ell_0}\right)^{-2}.
\end{equation}
The two autonomous relations, equations (\ref{lsdef}) and (\ref{bldef}),
are equivalent to  the self-similar relation eq.~(\ref{wdotm}) and the
assumption of gyrentropicity and can be used to replace self-similarity
when making a model of the inner disk (\cf \S~\ref{ssec:reldisks}). Note
that we are only free to specify two functional relations among $L,S,B$
because these relations must be consistent with the equations of motion,
equations (\ref{euler}) and (\ref{eulert}).

\subsection{Two-Dimensional Gyrentropic Disk Models}
\label{ssec:2dmod}

For an assumed value of the specific heat ratio, $\gamma$, the above relations
suffice to specify a two-parameter family of self-similar disk
models. To facilitate comparison with the one-dimensional models,
we choose these two additional parameters to be $R^{-1/2}L\rightarrow\ell_0$
and $RB\rightarrow b_0$. The mass loss exponent $n$ can then be
thought of as being given implicitly
by eq.~(\ref{bmidp}), which is identical to the 1D relation, eq.~(\ref{nlb}).
In addition, we must also specify the midplane entropy, $s_0$, which fixes the
pressure and density. However, $s_0$ does not change the shape
of either the disk or the isogyres and can be set to
unity without loss of generality.

\begin{figure*}
\includegraphics[width=18cm]{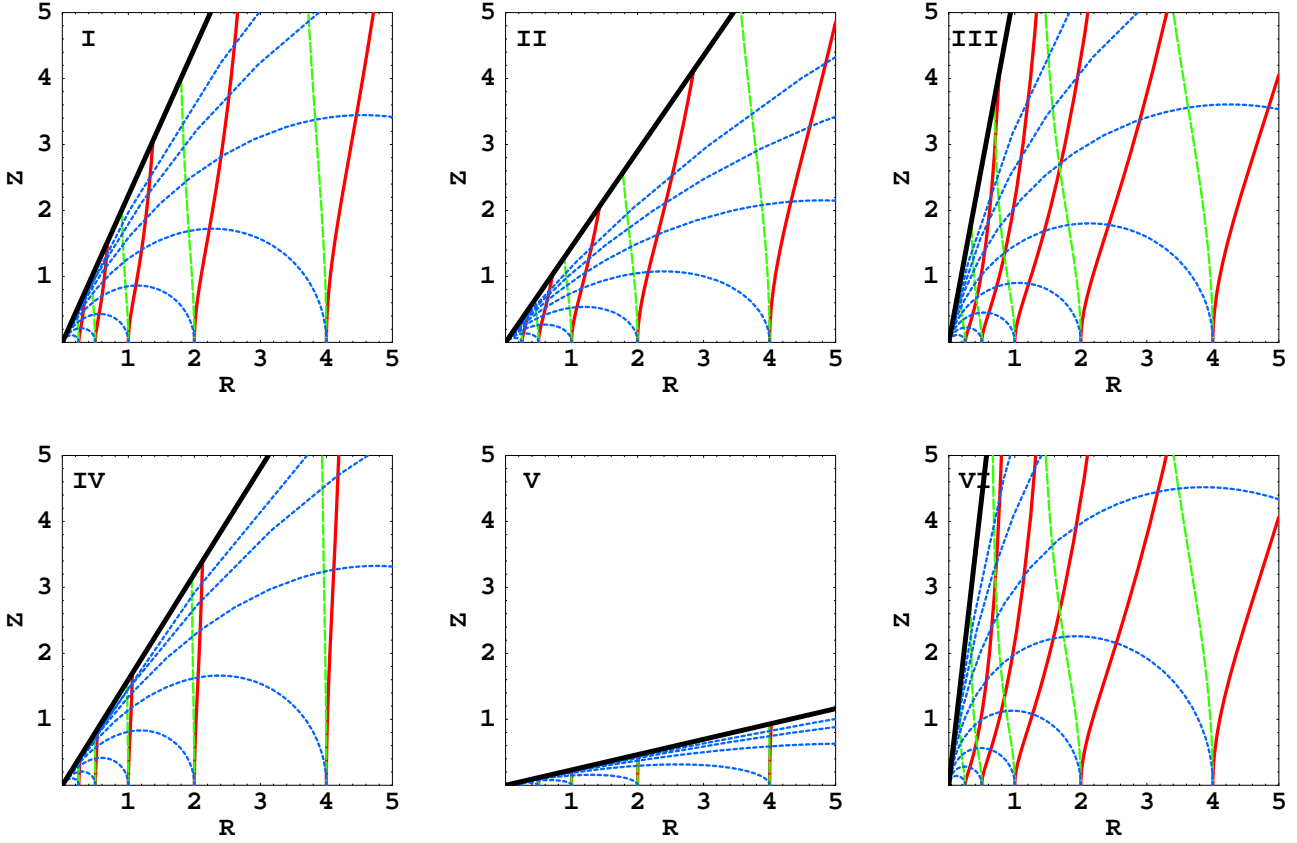}
\caption{The six simple 2D disk models described in
\S~\ref{ssec:2dmod} and Table~\ref{tbl:mod}.  The isobars, isorotes
and gyrentropes are shown with dotted, dashed and solid contours,
respectively. In these models, it is assumed that
convection is efficient and extends all the way to the disk surface.
The outflow does not influence the disk structure in these models.}
\label{fig:2d}
\end{figure*}

The six two-dimensional models, corresponding to the
six illustrative one-dimensional models (\S~\ref{ssec:1dmod}), are exhibited in
Fig.~\ref{fig:2d}. These models demonstrate
that the gyrentropes are generically negatively
curved surfaces, implying that the isorotes (surfaces of
constant $\Omega$) are positively curved.
They also show that the disks thicken, for a given equatorial angular momentum
$\ell_0$, as the mass loss rate, measured by $n$, decreases.
Conversely, for a given
value of $n$, the disks thicken as the rotational
support, measured by $\ell_0$, decreases and the pressure
becomes more important. Observe that
changing the specific heat ratio is relatively
unimportant. In this way, adiabatic accretion is rather different from
spherical accretion, where $\gamma=5/3$ constitutes a singular limit.

\section{Circulation Disks}
\label{sec:cir}

\subsection{Energy and Angular Momentum Transport}
\label{ssec:ene}

We have so far treated the disk as being in circular motion
and strictly gyrentropic.  We now relax the first
of these  assumptions.  Suppose that there is a local torque per unit length
\begin{equation}
\label{torden}
G\equiv{d\dot M{\cal G}\over ds}
\end{equation}
(where $ds$ is an element of length in the meridional plane),
that is small enough that we can treat its effect as perturbative
upon the fundamental,
gyrentropic flow pattern. (Note that $G$ is integrated over
azimuth.) We conjecture that this torque will eventually drive a
fluid disk to marginal instability everywhere,
so that the isentropes lie just ``inside'' the isogyres and that, except near
the disk surfaces, we can continue to approximate the disk angular
momentum distribution as gyrentropic.

As was discussed in \S~\ref{ssec:gyrent}, the convective motions
consist of slender ribbons of fluid moving between these two
surfaces. We suppose that the angle between the isogyres and isentropes
will open
up just enough to allow the relevant transport to take place. Heat
and angular momentum can be convected in this manner as a
displaced ribbon can exchange both quantities with its surroundings.
The relative efficiency with which this happens will depend upon
the effective Prandtl number, $Pr$. We assume that the
local thermal conductivity is much higher than the local effective viscosity,
implying that $Pr$ is large and that
the meridional flow is driven by the torque as opposed to
thermal conduction. It then follows that the displacements will nearly
follow the isogyres and that the convective
transport of heat $Q$ will be more important than the {\em convective}
transport of angular momentum, because the ribbons develop
larger departures from local thermal than mechanical
equilibrium.  We will consequently
ignore convective angular momentum transport and suppose that
the only torque is Newtonian, i.e., $G\propto-\nabla\Omega$.
(Alternative prescriptions are certainly possible and we shall consider
some in the following paper. However, we contend that our prescription is the
natural choice for high $Pr$, fluid disks.)

Convection, along gyrentropes, involves relatively rapid, forward
and backward motions that average to give a net mass flux
per unit length parallel to the gyrentropes, after integrating in azimuth.
To this must be added
the steady inflow (or outflow) of the gas perpendicular to the gyrentropes.
The two flows are combined into a single poloidal mass current vector
\begin{equation}
\label{masscur}
\vec J=2\pi R\Rho\vec V_p= r^{n-1}\vec j(\theta)
\end{equation}
where $\vec V_p\equiv[V_r,V_\theta]$.
\subsection{Torque Density}
\label{ssec:tor}
Consistent with the preceding discussion, we write
\begin{eqnarray}
\label{magtorq}
\vec G&\equiv&r^{n-1/2}\vec g(\theta)\\
&=&-2\pi\alpha R^3P\nabla\ln\Omega\\
&=&3\pi\alpha\sin^3\theta p[1,-2\omega'/3\omega]r^{n-1/2}
\end{eqnarray}
where $\omega=r^{3/2}\Omega=\ell\csc^2\theta$ is
the scaled angular velocity, $\alpha\equiv\nu\Rho\Omega/P$
is the standard viscosity parameter \citep{sha73} and $\nu$ is
the kinematic viscosity.
$\vec G$ can be treated as a poloidal vector in this approximation
although it is really a third rank tensor.
In order to comply with our
self-similar assumption, we set $\alpha=$ constant.
This ensures that the characteristic
inflow and circulation speeds are fixed fractions ($\propto\alpha$)
of the Keplerian value.

\subsection{Conservation Laws}
\label{ssec:conserv}

\begin{figure*}
\includegraphics[width=18cm]{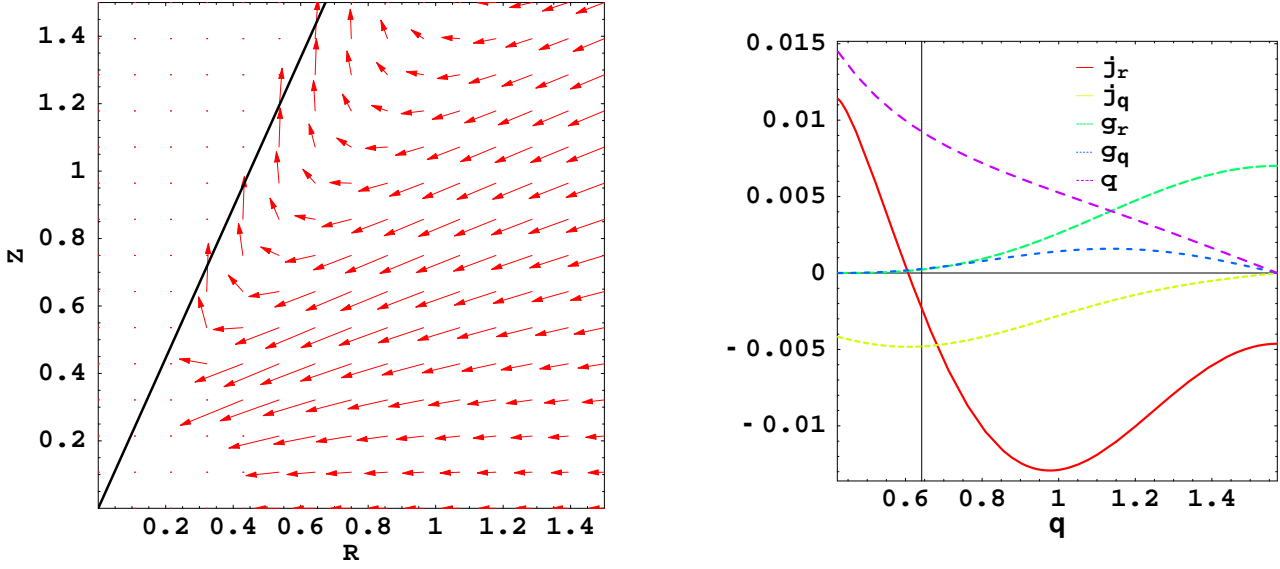}
\caption{Circulation model for the fiducial illustrative example I.
a) Poloidal mass flux $\vec j$ from eq.~(\ref{jode}). It is assumed that the
mass flow extends all the way to the disk surface. Note that,
although the net poloidal flow is inward, the flow is actually
outward near the disk surface.
b) The two components of the mass flux $\vec j$ and
the assumed viscous torque $\vec g$, as well as
the convective heat flow along the gyrentropes, measured by $q$. The vertical
line marks the location of the thermal front in the outflow model
(\cf \S~\ref{sec:out}).}
\label{fig:circ}
\end{figure*}

We can now write down equations representing the conservation
of mass and angular momentum:
\begin{equation}
\label{mlcons}
\nabla\cdot\vec J=0;\qquad\nabla\cdot(L\vec J+\vec G)=0.
\end{equation}
Adopting our self-similar prescription, these equations combine to give two
ordinary differential equations,
\begin{equation}
\label{jode}
j_\theta'
={2n\over\ell}[\ell'j_\theta+(n+{1\over2})g_r+g_\theta']=-nj_r .
\end{equation}
This can be solved for an assumed value of $n$ subject to the boundary
condition $j_\theta(\pi/2)=0$. Note that solutions to this equation
automatically conserve mass. If we start by assuming that $\dot M\propto
r^n$, then the net mass flowing across a hemispherical surface,
$\dot M=-r\int_{\theta_d}^{\pi/2}d\theta J_r$, automatically satisfies
$d\dot M/dr=-J_\theta(\theta_d)$, as required. The solutions for the
flow and the torque (Fig.~\ref{fig:circ}) show that there is a net
inflow across the gyrentropes and that
the total mass inflow at a given radius in a disk of given shape satisfies
$\dot M\propto s_0^{-\gamma/(\gamma-1)}\alpha$.

However, superimposed upon this inflow is a
quadrupolar circulation, directed inward in the equatorial region and
outward near
the disk surface. In some, though not all, solutions the
combined, radial flow is outward at high latitude.

Although the solution for $\vec J$ is finite at the disk surface,
the velocity $\vec V=\vec J/2\pi R\Rho$
diverges. This signals the failure of our model for
convective energy transport, as we discuss in the following section.
By contrast, the torque density $\vec G$ vanishes at the
disk surface according to eq.~(\ref{magtorq}).
The angular momentum incident on the disk surface is then
\begin{equation}
\label{surfam}
{dF_L\over dr}=-LJ_\theta=L{d\dot M\over dr}\Rightarrow\eta=0,
\end{equation}
where we have used eq.~(\ref{dfldef}). This confirms that
the outflowing gas only carries off
its own specific angular momentum, \cf eq.~(\ref{etalim}), BB99.

We next solve for the convective energy flux $\vec Q$.
This must be directed along the gyrentropes and so we can write it
in the form
\begin{equation}
\label{fcon}
\vec Q\equiv [2\ell',-\ell]\ q(\theta)r^{n-2} .
\end{equation}
The equation of energy conservation becomes
\begin{equation}
\label{encon}
\nabla\cdot[B\vec J+\Omega\vec G+\vec Q]=0.
\end{equation}
This furnishes a first order differential equation for $q(\theta)$,
\begin{equation}
\label{fode}
\ell q'-(2n-3)\ell'q+j_rb-j_\theta b'-(n-1)\omega g_r-\omega'g_\theta-
\omega g_\theta'=0,
\end{equation}
which can be solved subject to $q(\pi/2)=0$.

The supply of energy to the disk surface is given by
\begin{equation}
\label{surfen}
{dF_E\over dr}=J_\theta B+Q_\theta.
\end{equation}
Comparing with eq.~(\ref{dfedef}) we deduce that
\begin{equation}
\label{betadef}
\beta={Q_\theta\over J_\theta B}={\ell_d q_d\over j_{\theta d} b_d}=
{-2\sin\theta_dq_d\over j_{\theta d}}
\end{equation}
where $\ell_d,b_d$ are given by equations (\ref{ellsurf}) and
(\ref{conebern}) and
$j_{\theta d},q_d$ are given by
solutions of the differential equations
(\ref{jode}) and (\ref{fode}).
$\beta$ is then a function of $\ell_0$, $n$ and, implicitly
of $\ell_0$, $b_0$ as in the 1D models.

In the important limit when the disk expands
all the way to the pole, with $\theta_d,\ell_d\rightarrow0$ and
$\ell_0\rightarrow(2a)^{1/2},\ b_0\rightarrow1/2$, then $\beta\rightarrow
\infty$.
\subsection{Circulation Disk Models}
\label{ssec:circmod}
We have computed circulation disk models corresponding to the 1D, 2D
models discussed in \S~\ref{ssec:1dmod}, \S~\ref{ssec:2dmod}.
In Fig.~(\ref{fig:circ}), we exhibit the flow pattern for our fiducial
Model I (adopting a viscosity parameter $\alpha=0.03$).  We
also present the angular variations of the mass flux, torque and
convective heat flux.  The poloidal flow is fundamentally
quadrupolar, inward at low latitude and outward at high latitude.
The flows for the other five circulation
models are qualitatively similar.

\section{Outflow Disks}
\label{sec:out}
\subsection{Efficiency of Convection}
\label{ssec:eff}
We now turn to the second of our modifications to the basic
gyrentropic disk model, which introduces features
that may be largely absent from existing numerical simulations
due to inadequate resolution and/or missing physical processes.
We allow for the fact that convection of energy and angular momentum
will not be efficient close to the disk surface so that the
gyrentropic approximation may no longer hold.
We have already argued that the convection will
be along the isogyres. Assuming that the isogyres and the isentropes
have the unstable ordering with an angle $\delta$ between them, then,
according to standard mixing-length theory \citep[\eg][]{han94},
the convective energy flux, integrated over azimuth, will be given by
\begin{eqnarray}
\label{qcon}
Q&\sim&2\pi R\Rho v_c^3\\
&\sim&2\pi RP^{3/2}\Rho^{-1/2}(\vec h_P\cdot\nabla\ln S)^{3/2}\\
&\sim&2\pi RP^{3/2}\Rho^{-1/2}\delta^{3/2}
\end{eqnarray}
where $v_c$ is the convection speed and $\vec h_P$ is a vector of length
equal to the pressure scale height and tangent to the isogyre. (We drop
numerical constants of order unity.) The Mach number of the
convective motions is given by
\begin{equation}
\label{machcon}
M_c\sim\delta^{1/2}\sim\left({Q\Rho^{1/2}\over2\pi RP^{3/2}}\right)^{1/3}\sim
\left({q\ell\rho^{1/2}\over2\pi p^{3/2}\sin\theta}\right)^{1/3}.
\end{equation}

\subsection{Thermal Front}
\label{ssec:ths}

We now make the ansatz that, when $M_c$ reaches a certain
critical value $M_{c,crit}$, which we choose, arbitrarily,
though not unreasonably, to be unity,
the convective motions
rapidly dissipate and the convective energy flux is
quickly converted into heat, increasing the entropy of the gas.
Simultaneously, we suppose that the torque ceases to act
and that the viscous transport of angular momentum
and energy stops abruptly. In reality,
this transition is likely to occur gradually through the dissipation
of turbulent wave modes and the acceleration of the
flow within an extended region.  However, approximating this
transition as a discontinuity, located at a biconical
thermal front, suffices for us to make a simple model.
\begin{figure*}
\includegraphics[width=16cm]{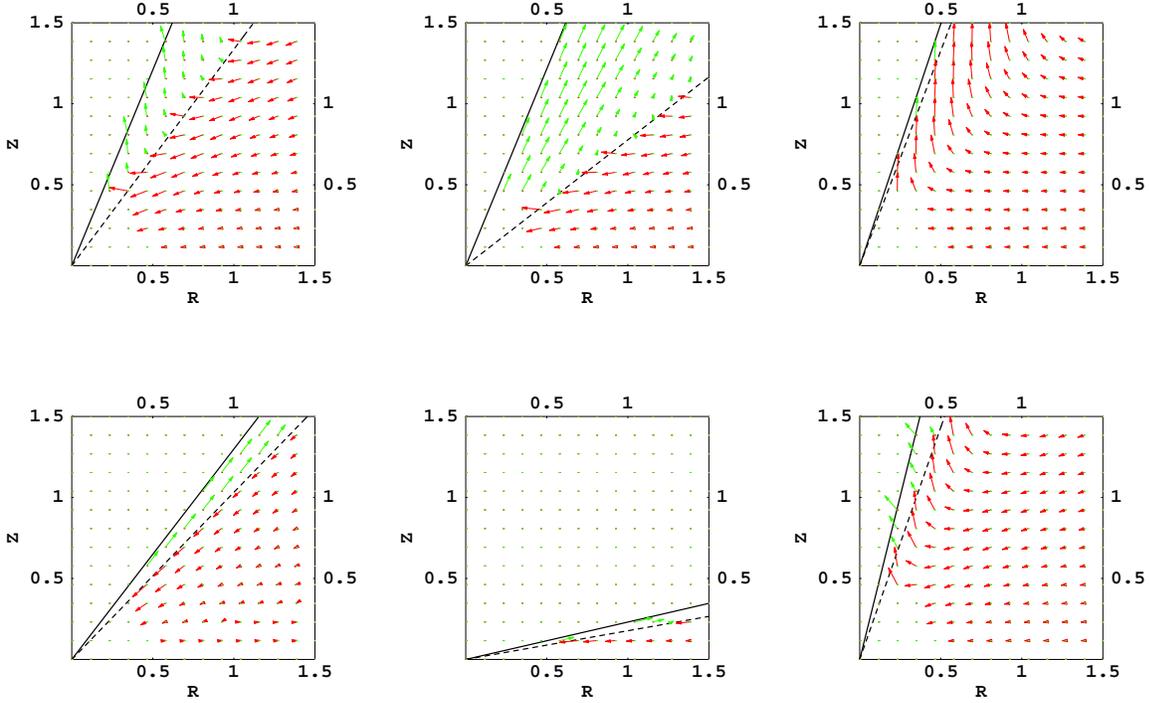}
\caption{Poloidal flows for the six outflow disk models
listed in \S~\ref{ssec:1dmod} and Table~\ref{tbl:mod}. Within the disk, the
conditions are identical to those presented in Fig.~\ref{fig:circ}.
However, the disk is presumed to terminate at a thermal front
(with opening angle $\theta_w$; dashed line) where the
convective heat flux and the viscosity abruptly vanish and across
which the flows of mass, momentum and energy are conserved. This is where
the outflow begins (\cf \S~\ref{ssec:ths}). The outflow lies between
cones with opening angles $\theta_j,\theta_w$. The flow at small radius is
omitted.}
\label{fig:out}
\end{figure*}

We impose jump conditions at this transition by
requiring that the mass, momentum and energy fluxes are equal
on either side of the transition.
\begin{eqnarray}
\label{thermmass}
[J_\theta]&=&0\Rightarrow j_{\theta w}=j_{\theta,th}\\
\label{thermmomr}
\left[J_\theta V_r\right]&=&0\Rightarrow v_{r,w}=v_{r,th}\\
\label{thermmomt}
\left[2\pi RP+J_{\theta}V_\theta\right]&=&0\nonumber\\
\Rightarrow j_{\theta,th}(v_{\theta,w}-v_{\theta,th})&=&2\pi\sin\theta_w
(p_{th}-p_w)\\
\label{thermmomp}
\left[G_\theta+J_\theta L\right]&=&0\Rightarrow
j_{\theta,th}(\ell_w-\ell_{th})=g_{\theta,th}\\
\label{thermen}
\left[G_\theta\Omega+J_\theta B+Q_\theta\right]&=&0\nonumber\\
\Rightarrow j_{\theta,th}(b_{th}-b_w)&=&
\ell_{th}q_{th}-g_{\theta,th}\omega_{th}.
\end{eqnarray}
These equations allow us to solve for the initial physical conditions
(specifically $p_w,\rho_w,\vec v_w$) at the
base of the wind (designated $w$) in terms of those upstream from the
thermal front (designated $th$). The disk now terminates at an opening angle
$\theta_w$ rather than $\theta_d$ and equations (\ref{surfam}) and
(\ref{betadef})
must be replaced with
\begin{eqnarray}
\label{etadefp}
\eta&=&{g_{\theta,th}\over\ell_{th}j_{\theta,th}}\\
\label{betadefp}
\beta&=&{\ell_{th} q_{th}-g_{\theta,th}\omega_{th}\over j_{\theta, th} b_{th}}.
\end{eqnarray}
Note that $\eta$ will now be negative because the isorotes
are less curved than spheres.
This is an artificial consequence of locating the effective disk surface
inside the disk. In general, $\eta$ is usually small for fluid disks and it is
a fair approximation to ignore it.

\subsection{Wind Solutions}
\label{ssec:outflo}

We now turn to the structure of a thermally-driven, gas dynamical wind
and describe simple similarity solutions
along the lines of those first derived by \citet{bar78}.
We suppose that a gas dynamical wind is launched from the surface of the disk
located at $\theta=\theta_w$ with initial conditions determined
by the jump conditions at the thermal front.  In the solutions that
we shall consider, the increase in entropy at
the thermal front is sufficient to change the sign of the Bernoulli
function and render the gas unbound. We suppose that in its subsequent
flow we can ignore viscous stress and mixing of mass, angular momentum
and energy between streamlines.

We continue to adopt the self-similar scalings and generalize the
velocity to a three-dimensional vector,
\begin{equation}
\label{windv}
\vec V=r^{-1/2}[v_r(\theta),v_\theta(\theta), v_\phi(\theta)],
\end{equation}
in spherical polar coordinates.
The equations of continuity, motion and entropy conservation can be written:
\begin{eqnarray}
\label{winda}
&&{\rho'\over\rho}+{v_\theta'\over v_\theta}=-\cot\theta-
{nv_r\over v_\theta}\\
\label{windb}
&&{v_r'\over v_r}={(v_r^2+2v_\theta^2+2v_\phi^2-2)\rho+(5-2n)p
\over2\rho v_rv_\theta}\\
\label{windc}
&&{v_\theta'\over v_\theta}+\left({p\over\rho v_\theta^2}\right){p'\over p}=
{2v_\phi^2\cot\theta-v_rv_\theta\over2v_\theta^2}\\
\label{windd}
&&{v_\phi'\over v_\phi}=-{v_r\over2v_\theta}-\cot\theta\\
\label{winde}
&&{1\over\gamma}{p'\over p}-{\rho'\over \rho}=\left({a\over1-a}
\right){v_r\over v_\theta}
\end{eqnarray}
where a prime denotes differentiation with respect to $\theta$.

These equations can now be rearranged:
\begin{eqnarray}
\label{windf}
p'&=&\gamma p\rho[(1+2n(a-1)+a)v_rv_\theta\nonumber\\
&&{\qquad\qquad+2(a-1)(v_\phi^2+v_\theta^2)\cot\theta]\over
\qquad2(a-1)(\gamma p-\rho v_\theta^2)}\\
\label{windg}
\rho'&=&-\rho[v_r(2\gamma pa +(2n-1)(a-1)\rho v_\theta^2)\nonumber\\
&&{\qquad\qquad+2(a-1)\rho v_\theta(v_\phi^2+v_\theta^2)\cot\theta]\over
\qquad2(1-a)v_\theta(\gamma p-\rho v_\theta^2)}\\
\label{windh}
v_r'&=&{(5-2n)p+\rho(2v_\phi^2+v_r^2+2v_\theta^2-2)\over2\rho v_\theta}\\
\label{windi}
v_\theta'&=&[v_r(-2\gamma p(a+n a-n)+(a-1)\rho v_\theta^2)\nonumber\\
&&{\qquad\qquad-2(a-1)(\gamma p+\rho v_\phi^2)v_\theta\cot\theta ]
          \over\qquad2(a-1)(\gamma p-\rho v_\theta^2)}\\
\label{windj}
v_\phi'&=&{-v_\phi(v_r+2v_\theta\cot\theta)\over2v_\theta} \ .
\end{eqnarray}
These equations have three integrals, as may be readily verified.
The entropy $S$, the Bernoulli function
$B$ and the specific angular momentum $L$ are conserved along streamlines.
We conclude that, as with the disk, $S=S(L),B=B(L)$ in the wind,
although for a quite different reason. The entire flow can then be
labeled using the disk gyrentropes that extend out to the thermal front
where they are replaced by the wind gyrentropes.

Now consider the wind flow well away from the disk. There is potentially
a critical point
where $v_\theta=(\gamma p/\rho)^{1/2}$, the adiabatic sound speed. (The reason
why only the $\theta$ component of the velocity is involved is
that self-similarity and axisymmetry prescribe the other
two components.)  However, solutions satisfying appropriate boundary
conditions never reach their critical speeds.
 From a mathematical point of view, these winds are
``breezes," and $v_\theta$ is always subsonic, even if the
radial velocity becomes strongly supersonic. (In fact, $v_\theta$
must be subsonic, under all conditions, for $\gamma<3/2$.)

As $B$ is conserved in the outflow we can use its value at the thermal front
to compute the asymptotic wind speed.  We express the speed, $v_\infty$,
by tracing the flow back to the thermal front on a flow line
and then back to the midplane on a gyrentrope. $v_\infty$ is then
given in Table~\ref{tbl:mod} in units of the Keplerian speed at this
point.

\subsection{Outflow Disk Models}
\label{ssec:outmod}

We have integrated equations (\ref{windf})--(\ref{windj}) and produced
matched flows for
the six examples (Fig.~\ref{fig:out}). We find that it is generally possible
to obtain physically plausible solutions
for viable disk models. For each of our examples,
which span a much larger volume of parameter space than we envisage
is occupied by real disks, we find that the viscous dissipation creates
so much heat that it easily unbinds the gas downwind of the thermal front.
The outflow always occupies a hollow cone
on account of the centrifugal barrier. The
computed thermal fronts lie somewhat below the surfaces of the computed
2D disks (\ie, $\theta_w>\theta_d$) and so the outflow disks
are somewhat thinner
than the equivalent circulation disks, but not by a large factor
for reasonable values of $\alpha$ and $M_{c,crit}$.
The jet cones have angles $\theta_j$ that are
not much smaller than the thermal front cone angles, $\theta_w$.
The values of $\beta$ computed in these more complete models do not differ
greatly from those in our simple 1D and circulation disk models. The values
of the angular momentum loss parameter $\eta$ are generally small
except for Model IV, where it is compensating for the low mass loss rate.

We have made a very simple, two-part model to account for the
dissipation. We suppose that there is distributed, Newtonian dissipation
within the disk, and a discontinuous entropy production front at the
disk surface. How do our results depend upon the choice of $\alpha$,
which controls the former process, and $M_{c,crit}$, which controls the latter?
For Model I, we find that increasing $\alpha$ from 0.03
to 0.1 causes the thermal front to be located at greater density in the disk,
where the opening angle is $\theta_w=0.78$ as opposed to 0.64.
The cone excluded by the jet has a correspondingly larger opening
angle, increasing
to $\theta_j=0.46$ from 0.39. Conversely, when we reduce $\alpha$ to
0.01, the thermal front is located
at $\theta_w=0.57$, i.e., closer to the surface of the original
gyrentropic disk at $\theta_d=0.42$,
so that the original gyrentropic model becomes more accurate. (In
addition, $\theta_j$ falls to 0.33.)

When we separately reduce $M_{c,crit}$ to 0.7 we find that the thermal front
is located deeper in the disk, at $\theta_w=0.76$,
and that the jet cone angle increases to  $\theta_j=0.55$ as the outflow is
launched with a lower speed relative to the surface of the thermal front.
Our models are therefore not strongly sensitive to the values of
$\alpha$ and $M_{c,ccrit}$, and the
sense of the changes that variations in these parameters bring about
are as expected.

\section{Non-self-similar Flow at Small and Large Radii}
\label{sec:nss}

\subsection{Energy Release in Adiabatic Flows}
\label{ssec:lims}

We have so far concentrated upon self-similar models of adiabatic
accretion flows because these allow us to generate self-consistent solutions
through solving ordinary differential equations.  However, as was
made clear in BB99, the requirement that energy be carried away
somehow from an adiabatic flow follows from general conservation laws
and is not dependent upon
self-similarity. If, for example, the viscous torque were to decrease
suddenly at the radius where the disk becomes neutral, the thickness and the
outflow would also be expected to exhibit abrupt changes.
However there would still be a need for the energy released at small
radius to be carried off in an outflow. In the context of the present
paper,
the disks should still convect energy to their surfaces and
drive meridional circulation.

Furthermore, self-similar solutions cannot describe
important features of the flow at small and large radius where self-similarity
must fail. In this section, we discuss relativistic
flow at small radius and non-self-similar flow near the
outer transition radius.

\subsection{Relativistic Flow at Small Radius}
\label{ssec:reldisks}
A convenient form of the equation of hydrostatic balance for a
stationary, axisymmetric flow is
\begin{equation}
\label{relhydros}
-{\nabla P\over\tilde\Rho\tilde H}=\vec{\tilde A}=
\nabla\ln\tilde E-{\tilde\Omega\nabla\tilde L
\over1-\tilde\Omega\tilde L}
\end{equation}
where $c=1$, $\vec{\tilde A}$ is the acceleration
and the relativistic enthalpy per unit rest mass is
$\tilde H=1+H=1+\gamma P/(\gamma-1)\tilde\Rho$, with $\tilde\Rho$
now representing the proper density of rest mass.
This applies to both radiation-dominated ($\gamma=4/3$) and ion-dominated
($\gamma=5/3$) flows.
$\tilde E=-u_0$ is the energy at infinity, $\tilde\Omega=u^\phi/u^0$
is the angular velocity and $\tilde L=-u_\phi/u_0$ is the fluid
angular momentum. [Note that the relativistic definition of angular momentum
differs from the conventional dynamical choice of $u_\phi$. It is more
convenient for fluid dynamical use \citep{bar73,seg75}.]
These essentially kinematical quantities
are related through
\begin{eqnarray}
\label{kerrmet}
\tilde E(r,\theta,\tilde L)&=&(-g^{00}+2g^{0\phi}\tilde L-g^{\phi\phi}
\tilde L^2)^{-1/2}\cr
\tilde\Omega(r,\theta,\tilde L)&=&{g^{0\phi}-g^{\phi\phi}\tilde 
L\over g^{00}-g^{0\phi}
\tilde L} .
\end{eqnarray}
where $g^{\alpha\beta}$ is the contravariant form of the metric tensor in
Boyer-Lindquist coordinates for a hole with specific angular momentum $a$.
For illustration we adopt the value $a=0.95m$.

The relativistic generalization of
eq.~(\ref{gradbern}) is
\begin{equation}
\label{relbern}
\nabla\ln\tilde B=(1-\tilde H^{-1})\nabla\ln S+
{\tilde\Omega\nabla\tilde L\over1-\tilde\Omega\tilde L} ,
\end{equation}
where
\begin{equation}
\label{relberndef}
\tilde B=\tilde H\tilde E
\end{equation}
is the relativistic Bernoulli function and the entropy function
is still given by eq.~(\ref{entdef}). Note that the Bernoulli
function now includes a contribution from the rest mass.

The two non-relativistic H\o iland criteria, equations (\ref{hoione}) 
and (\ref{hoitwo}),
have relativistic counterparts \citep{seg75,bla85}. The first 
criterion for marginal instability generalizes to become:
\begin{equation}
\label{relhoione}
{1\over\tilde H^2}\nabla\left({P^{1-1/\gamma}\over1-1/\gamma}\right)
\cdot\nabla S -\vec\gamma\cdot\nabla\tilde L>0
\end{equation}
where
\begin{equation}
\label{gammadef}
\vec\gamma={d\vec {\tilde A}\over d\tilde L}=
{E^4\nabla\tilde L\over g_{0\phi}^2-g_{00}g_{\phi\phi}}-
{\nabla\tilde\Omega\over(1-\tilde\Omega\tilde L)^2}
\end{equation}
and the derivative is performed assuming only circular motion.
The second criterion also generalizes to become
\begin{equation}
\label{relhoitwo}
(\nabla P\times\vec\gamma)\cdot(\nabla S\times\nabla\tilde L)>0.
\end{equation}
We find that the first H\o iland criterion is even more strongly satisfied
in the inner disk (where the entropy gradient changes to become stabilizing)
than in the self-similar disk.
We conjecture that the flow will evolve to a state of marginal
stability according to the second H\o iland criterion
so that the entire disk is gyrentropic.

\begin{figure*}
\includegraphics[width=16cm]{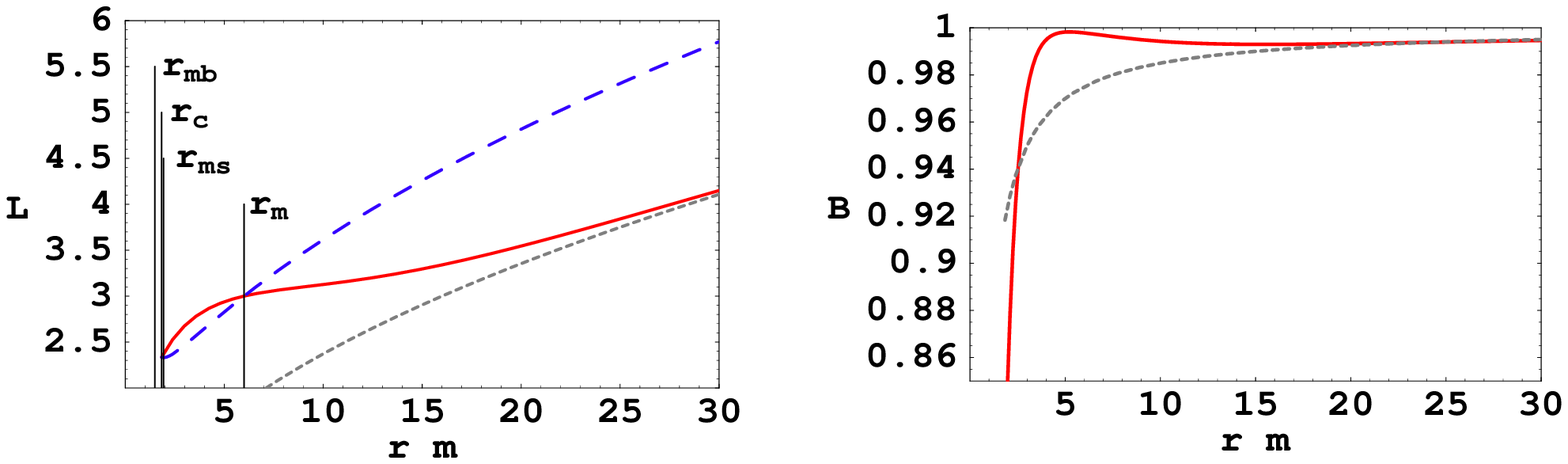}
\caption{a) Assumed relativistic, fluid angular momentum variation,
$\tilde L$ (solid) for a gyrentropic disk around a Kerr hole
with $a=0.9m$. The Keplerian variation (dashed) is shown for comparison.
The two angular momenta must coincide at the cusp radius $r_c=1.84m$ and at the
pressure maximum at $r_m=6m$. At large radius the fluid angular momentum
asymptotes to the Model I self-similar solution (dotted). 
Also shown are the radius
of zero binding energy $r_{mb}=1.5m$ and the radius of marginal stability
$r_{ms}=1.93m$.
b) Derived variation of the relativistic Bernoulli function
$\tilde B(r_0)$ (dashed). Like the entropy, it asymptotes quickly to the 
non-relativistic form $B(r_0)=b_0r_0^{-1}$.}
\label{fig:gyr}
\end{figure*}

In order to construct a gyrentropic inner disk model,
we need to introduce two autonomous functions $S(\tilde L),\tilde B(\tilde L)$
as described in \S~\ref{ssec:gyr}. What we actually
do is ultimately equivalent.  We specify equatorial distributions
of angular momentum and entropy, $\tilde L(r_0),\tilde S(r_0)$,
from which we can deduce $\tilde S(\tilde L)$,
that are chosen to match onto the non-relativistic
functions at large radius. In addition,
the angular momentum must equal the
relativistic Keplerian value at two radii --- a radius $r_m$
where the pressure is maximized
and a smaller radius $r_c$ (the ``cusp'' radius) where the pressure
and its gradient vanish.  
The particular form of $\tilde L$ that we use is
\begin{equation}
\label{relangmom}
\tilde L(r_0)=\ell_0r_0^{1/2}[1+c_1\exp(- c_2r_0)]
\end{equation}
where the constants $c_1=1.78,$, $c_2=0.18$ 
are chosen to fit a disk with $r_c=1.84 m$, $r_m=6m$ (Fig.~\ref{fig:gyr}).

The entropy function is required to
satisfy the first H\o iland criterion and has to be tuned to allow the 
Bernoulli function, derived below, to match the 
non-relativistic form $B(r_0)=b_0r_0^{-1}$ at large radius. We use
\begin{equation}
\label{relentropy}
S(r_0)=r_0^{-a\over1-a}[1+c_3\exp(-c_4 r_0)] .
\end{equation}
The choices $c_3=5.0,c_4=0.85$ produce a suitable solution.

In order to compute the Bernoulli function, we use eq.~(\ref{kerrmet})
to compute the energy $\tilde E$ and the angular velocity $\tilde\Omega$
as functions of equatorial radius $r_0$.
We then rewrite eq.~(\ref{relbern}) in the form
\begin{equation}
\label{relbernr}
{d\tilde B\over dr_0}=(\tilde B-\tilde E(r_0))
{d\ln S\over dr_0}(r_0)+{\tilde\Omega(r_0)\tilde B\over1-
\Omega(r_0)\tilde L(r_0)}{d\tilde L\over dr_0}
\end{equation}
and solve this differential equation to infer the variation
$\tilde B(r_0)$ and, consequently, $\tilde B(\tilde L)$ (Fig.~\ref{fig:gyr}).

\begin{figure*}
\includegraphics[width=16cm]{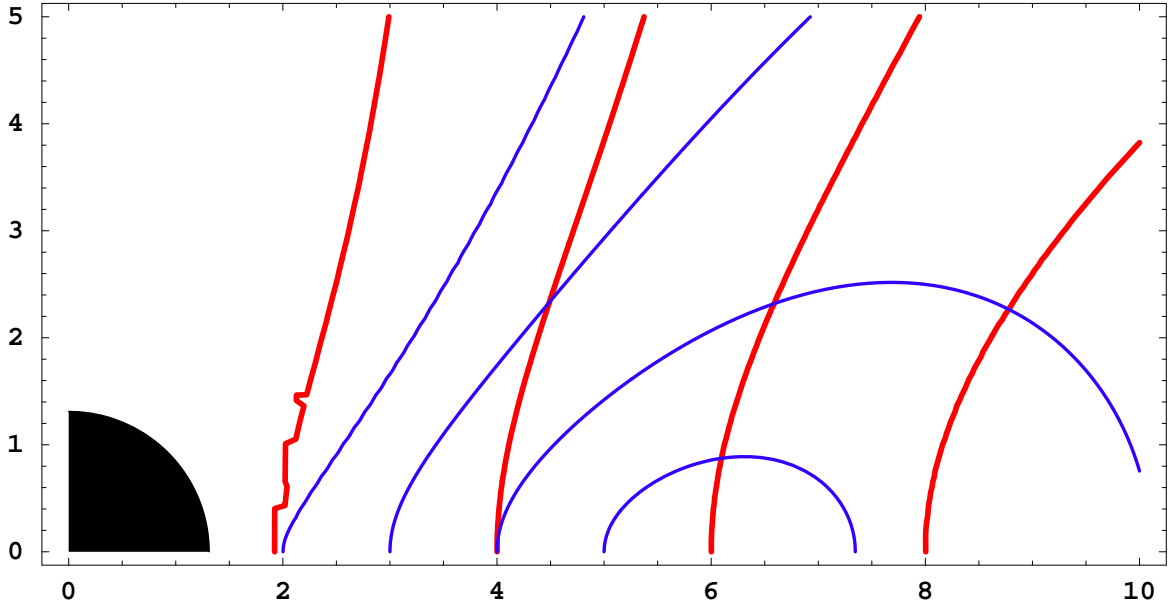}
\caption{Model of the inner regions of  gyrentropic accretion 
disk orbiting a Kerr black
hole with the assumed angular momentum and entropy distributions displayed in
Fig.~\ref{fig:gyr}.
The gyrentropes are displayed as bold
contours. The cusp is located between the marginally bound and marginally
stable orbits.
The isobars are shown as light contours surrounding the pressure maximum.}
\label{fig:tor}
\end{figure*}

The next step is to solve for the disk structure away from the equatorial
plane. Again we use eq.~(\ref{relbern}), this time written in the form:
\begin{equation}
\label{relbernl}
{d\tilde B\over d\tilde L}(\tilde L)=(\tilde B(\tilde L)-
\tilde E(r,\theta,\tilde L))
{d\ln S\over d\tilde L}(\tilde L)+{\tilde\Omega(r,\theta,\tilde L)
\tilde B(\tilde L)\over1-
\Omega(r,\theta,\tilde L)\tilde L}
\end{equation}
and solve this for $\tilde L(r,\theta)$. This enables us to calculate
$S(r,\theta),B(r,\theta)$ and then to compute
$E(r,\theta),\Omega(r,\theta)$ using eq.~(\ref{kerrmet}).
Finally, we can calculate the pressure using
\begin{equation}
\label{press}
P=\left[\left({\gamma-1\over\gamma S}\right)
\left({\tilde B-\tilde E\over\tilde E}\right )\right]^{\gamma\over\gamma-1}
\end{equation}
and plot the isobars (Fig.~\ref{fig:tor}).

Although our
choice of functions in equations (\ref{relangmom}) and (\ref{relentropy}) 
is arbitrary and has no immediate
physical basis, our procedure does illustrate how to construct
self-consistent disk models given a more comprehensive
theory of relativistic convection.
It is straightforward, in principle, to generalize the non-relativistic,
self-similar development of circulation and outflow disk models to the
non-self-similar regime. We could use the relativistic generalizations of
equations (\ref{mlcons}) and (\ref{encon})
to derive partial differential equations for
the poloidal velocity for a non-self-similar circulation and outflow. 
However, there is
little point in carrying out this exercise as the assumed
autonomous relations that define the gyrentropes are arbitrarily prescribed.
It is, instead, more valuable to solve the time-dependent
2D and 3D fluid dynamical
equations numerically, making different assumptions about the
viscosity and, perhaps, the heat transport. The
principles that we have developed here should be of use in interpreting
such a computation.

\subsection{Transition Disk}
\label{ssec:trans}

Self-similarity must also break down at some outer radius, which
either represents an outer boundary condition associated with the
mass supply or the radius beyond which radiative losses are
significant. In the ion-dominated case, cooling near the outer
transition radius $r_t$ (i.e., between a thin radiative disk and an
ADIOS) is likely to be dominated by thermal bremsstrahlung of the
one-temperature ($T_e = T_i$) plasma, provided that $r_t > 2000 m$
(where $m = GM/c^2$ is the gravitational radius).  Equating the
cooling rate and the inflow time, we find that adiabatic flow is
possible for
\begin{equation}
\label{rtrans}
r_t <  10^9 {\alpha^4 \over \dot m_t^2} m,
\end{equation}
where $\dot m_t \equiv \dot M_t / \dot M_E$ is the mass flux crossing
into the nonradiative region (the mass supply).  Given a power-law
scaling of $\dot m$ with index $n$ at $r< r_t$, the mass flux
reaching the black hole (the accretion rate) is
\begin{equation}
\label{rtransm}
\dot m_{\rm acc} <  10^{-9n} \alpha^{-4n} \dot m_t^{1+ 2n}.
\end{equation}
For example, if $\dot m_t = 0.01$, $\alpha = 0.03$ and $n=0.6$ (as in
Model I), the accretion rate can be as low as $\dot m_{\rm acc}\sim
10^{-6}$, four orders of magnitude lower than the supply rate.  Since
$\dot m_{\rm acc} \ll 50 \alpha^2 \sim 0.05$, the ADIOS easily
satisfies the condition for a two-temperature flow close to the black
hole \citep{ree82} --- which is more stringent than the condition for
adiabatic flow at large radii --- implying that the adiabatic flow,
once established, will extend all the way from $r_t$ in the vicinity of
the hole.  For conservative inflow models, this condition is more
difficult to satisfy. Moreover, for the parameters we have chosen
$r_t$ can be as large as $\sim 10^7 m$; thus the flow could exhibit
self-similarity over several decades in radius.

Similar considerations apply to accretion flows with $\dot m \gg 1$.
Here the transition from radiative to nonradiative flow occurs at the
trapping radius, $r_t \sim r_{\rm trap}\sim \dot m_t m$ 
\citep{beg79}. The accretion rate is given by $\dot m_{\rm acc}\sim 
\dot m_t^{1-n}$.

We note that models for the transition between the thin disk and
nonradiative flow have been studied mainly in the context of ADAF
models, in which conserved mass flux through the disk is assumed
\citep[\eg][]{kat98,man00}.  These models generally require a zone
of anomalous emissivity near $r_t$ in order to soak up the outward
energy flux, which cannot be accepted by the thin disk in a smooth
transition.  This problem is avoided if we allow for the onset of mass
(and angular momentum) loss in the transition region.

In order to demonstrate that a smooth transition is possible under ADIOS
models, we combine the analyses of \S~\ref{ssec:cons} and 
\S~\ref{ssec:reldisks}
and make a one-dimensional model for a disk where we choose 
functional forms for the radial variation
of $L,B,{\cal G}$ that interpolate between the thick and thin limits. We
emphasize that these functions, although plausible, have no physical basis
and are only intended to demonstrate how a physical solution might be
constructed. In particular, these models are not consistent with a 
simple $\alpha-$type parametrisation of the viscous couple, in which 
${\cal G}$ is an increasing function of pressure. It may not be 
possible to construct self-consistent models for all reasonable forms 
of ${\cal G}$; in some cases, solutions may be time-dependent.

We choose to match to model I and adopt a self-similar
thin disk with $\ell_{\rm out}=0.95,b_{\rm out}=-0.45$. We require that the
mass and angular momentum loss rates $\rightarrow0$ as 
$R\rightarrow\infty$ and that the
energy loss rate approach the standard value for a thin
disk $\ell_{\rm out}^2-b_{\rm out}$ per unit mass flow. We anticipate 
that this energy
loss will change from outflow to radiation with increasing radius, though
we do not need to specify how rapidly this occurs. We do require the 
flows of mass,
momentum and energy, $\dot M$, $F_L$, and $F_E$, to vary monotonically
with radius; this does restrict the types of function that we choose.
(Note that there is a constant positive angular momentum flux, $F_{L, {\rm 
out}}$, flowing
inward through the outer disk.) However there is still considerable
freedom left and so we do not believe that these solutions are 
particularly contrived.

Without loss of generality we set the mass supply rate from the
outer disk and the pressure at $R=1$, in the middle of the transition
region, to unity. We also simplify matters by setting $\eta=0$.
Our interpolating functions are:
\begin{eqnarray}
\label{transfn}
L(R)&=&{\{\ell_{\rm out}+\ell_0
+(\ell_{\rm out}-\ell_0)
\tanh[\ln R/w]\}R^{1/2}\over2}\\
B(R)&=&{b_{\rm out}+b_0
+(b_{\rm out}-b_0)
\tanh[\ln R/w]\over2R}\\
F_L(R)&=&{2n\ell_0R^{1/2}\dot M\over(1+2n)
\left[1+\left({2n\ell_0\over(1+2n)F_{L, {\rm out}}}\right)^2R\right]^{1/2}}
\end{eqnarray}
and ${\cal G}$ is given by eq.~(\ref{fldef}). After some experimentation,
we find that the choices $w=0.8,F_{L, {\rm out}}=0.5$ lead to plausible
solutions. It is straightforward to solve for the sound speed,
pressure, density and entropy function using eq.~(\ref{berndef})
and integrating the radial equation of motion.
We then solve for $\dot M$ by integrating eq.~(\ref{dfldef}) and $F_E(R)$ using
eq.~({\ref{fedef}). Our results for this particular example are
exhibited in Fig.~\ref{fig:tra}.
\begin{figure*}
\includegraphics[width=16cm]{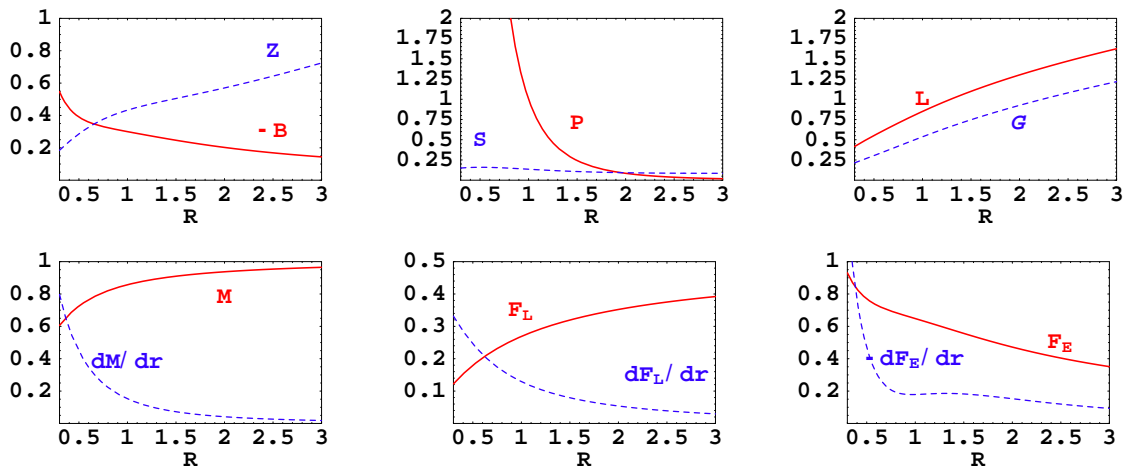}
\caption{Model of the transition disk that connects the self-similar
disk to an outer thin disk. The inner disk is chosen to have the parameters
of fiducial Model I. The outer disk is a much thinner self-similar disk
with $L(R)=0.95R^{1/2}$, $B(R)=-0.45R^{-1}$. The top three panels show
the variation of the Bernoulli function $B$ and the disk thickness
$Z=(P/\Rho\Omega^2)^{1/2}$, the pressure $P$ and the entropy $S$,
the angular momentum $L$ and the torque ${\cal G}$. The bottom three panels
show the flows of mass, angular momentum and energy through the disk, and
the associated outflows per unit radius.}
\label{fig:tra}
\end{figure*}

These one-dimensional results can be used to verify that the disk remains
stable according to the first H\o iland criterion. If the disk remains
gyrentropic, and the case for this weakens as it becomes increasingly 
radiative,
then it is straightforward to repeat the analysis carried out in
\S~\ref{ssec:reldisks} and construct a two-dimensional model. The gyrentropes
remain close to right cylinders in the transition disk.

\subsection{Choice of Disk--Outflow Model}
\label{ssec:choice}
We argued in \S~\ref{ssec:cons} that three internal parameters that 
characterise
self-similar disks, which we chose to be $\ell_0$, $b_0$ and a torque
parameter, determine the nature of the outflow, as measured
by $n$, $\beta$ and $\eta$. How is this linkage established in practice?
A comprehensive answer to this question must await more numerical experiments
but some qualitative guidelines can be uncovered using our models. 
The first point is that angular momentum and mass are supplied to the 
adiabatic flow
at the transition radius and are lost as the gas flows inward. 
Likewise, the energy derives from the relativistic
regime and flows outward. The ratio of angular momentum flux to 
supplied mass flux in the transition region, $F_L/\dot M$, which is 
presumably set by local and outer boundary conditions, can have an 
influence on the subsequent inflow. If this ratio is large, then the 
disk will rotate rapidly and this can be propagated inward. Likewise, 
the energy per unit accreted mass --- $F_E/\dot M$ --- emerging from 
the relativistic inner disk controls the
disk pressure and thickness and this can propagate outward.

However, this is not the whole story.  The local physics within the
self-similar regime is also important.  If the rate of production
of entropy near the disk surface is high, then mass loss will 
increase. Conversely, when it is low there will be more internal
circulation to larger radii, where energy can escape more easily. If 
the disk has an organized magnetic field, then the loss of angular 
momentum per unit
mass will be larger and the internal angular momentum transport will be
diminished. Mathematically, we should think of coupled differential 
equations for the flows of mass, energy and angular momentum that 
have to be
solved with boundary conditions at both small and large radius,
just as with the theory of stellar structure.

\section{Alternative Models of Adiabatic, Fluid Accretion}
\label{sec:alt}
\subsection{Non-Convective Models}
\label{ssec:noncon}
In an alternative description of thick accretion disks, pioneered by
\citet{pac82} for radiation-dominated tori,
(\cf \citet{ree82} for the ion-dominated case),
the flow is idealized as a quasi-stationary torus orbiting the
black hole, in hydrostatic equilibrium with an {\it ad hoc} entropy
distribution that is commonly chosen to be barytropic, $P=P(S)$.
If this flow were convectively stable, the entropy would have to
rise vertically as
there can be no rotational stabilisation when a H\o iland interchange
is performed in the vertical direction.
As there is a pressure maximum, and the isentropes coincide with the
isobars in a barytropic flow, the isentropes would have to be
a set of nested tori with minimum entropy at the pressure maximum.

However, this configuration cannot be stationary. To see this, note
that the conservation laws, equations (\ref{mlcons}) and (\ref{encon})
(with $Q=0$), combine to give the entropy conservation law
\begin{equation}
\label{entrocon}
\vec G\cdot\nabla\Omega+H\vec J\cdot\nabla\ln S=0 .
\end{equation}
Now, with a local torque, the first term
in eq.~(\ref{entrocon}) will be negative, consistent with the second law of
thermodynamics (\cf \citet{lan59}).
This implies that $\vec J \cdot \nabla \ln S > 0$, i.e.,
there must be a mass flux directed toward increasing entropy.
However, this would mean that there must be a mass
source ($\nabla\cdot \vec J \neq 0$) at the pressure maximum. It is
therefore clear
that convectively stable, barytropic tori cannot be stationary.
This conclusion is not altered by the inclusion of general relativity
and is probably generally true for two-dimensional, H\o iland-stable
disk flows, although we have not been able to give a proof.
Typically, what will happen is that any initially stable
flow will quickly become convective, due to the action of the viscous stress,
and the topology of the isentropes will change so that the entropy
decreases with increasing cylindrical radius.

\subsection{Advection-Dominated Accretion Flows}
\label{ssec:adaf}

As outlined in \S~\ref{sec:int}, the original ADAF idea
\citep{ich77,nar94,nar95}
was that the inflow would be
conservative with all the mass supplied at large radius flowing across
the horizon. This would probably be the case as long as there were no
angular momentum.  However, as soon as the inflow is
controlled by viscous torque that
can be treated perturbatively, there seems no escape from the conclusion
that most of the mass supplied
escapes in an outflow, powered by the energy released by the small fraction
of the gas that accretes onto the hole. (It is, in principle, possible
that the viscous torques could be strong enough to ensure radial
inflow which would vitiate this conclusion,
but there is strong analytical and numerical evidence that neither
fluid nor magnetofluid torques could ever be this large.)
At a global level, the physical inconsistency is manifested
in a positive Bernoulli function, indicating that parcels of
matter near the surface of the disk can escape to infinity with
positive energy after doing work on their surroundings.  At a local
level, the positive energy condition implies convective instability
(in the hydrodynamic limit), providing a framework for understanding
the mechanism of energy, angular momentum, and mass transport. In Paper II,
we will show that analogous arguments carry over to magnetohydrodynamic
disks, despite the fact that the stability and transport mechanisms
are likely to
be quite different.

For these reasons, we argue that black holes surrounded by radiatively
inefficient flows are likely to accrete matter at a rate far smaller
than the rate at which mass is supplied at large radii.  The gas density
close to the hole, from where most of the observed emission derives, is
therefore likely to be orders of magnitude
smaller than derived under the ADAF framework, implying that interpretations
of observational data within the latter framework are
incorrect. We note, however, that the ADIOS framework developed in this
paper does not automatically predict the relationship between the
accretion rate and the supply rate.  This depends not only on additional
information about the outer supply (or ``transition") radius, but also
on the microphysical mechanisms that increase the entropy and power the
wind at the thermal front. (In particular, the MHD winds to be discussed in
Paper II may be capable of carrying away the required energy and
angular momentum with  very little loss of mass.)

\subsection{Convection-Dominated Accretion Flows}
\label{ssec:cdaf}

The CDAF idea was developed to address the energetic difficulties
encountered by ADAFs \citep{nar00,qua00b}.
These models are based on the surprising fact that fluid dynamical
turbulence, as might arise, for example, from convection, is capable
of transporting angular momentum radially {\it inward} in an accretion
disk \citep[\cf][]{ryu92,bal00,qua00a}.  In CDAF models, it is supposed
that the inward convective
stress essentially balances the outward viscous stress locally.
As a consequence, both the mass flux and the net angular momentum flux,
which have different natural scalings with radius, must vanish while
the energy flux is directed outward.  In the two-dimensional model
of \citet{qua00b}, $B\sim0$ everywhere and the flow must
extend all the way to the pole.   More detailed models assume that $B$
is very small, of order the binding energy at the outer (or transition)
radius, and that a small conservative accretion flow is responsible for
powering the outward flux of energy. The latter is assumed to be disposed
of (e.g., by radiation or slow outflow) far from the black hole.

As convective transport is essentially nonlocal, there is no
thermodynamic objection
to the average convective torque being negative, although there may
be thermodynamic limitations on the extent to which the (macroscopic)
convective stress can cancel the (microscopic) viscous stress locally
\citep{bal02}. However, the invocation of radial convective
transport does seem arbitrary, given the nature of the convective
instability. Recall that, at marginal stability in the hydrodynamic
(non-MHD) limit, the
unstable motions are along the gyrentropes, along roughly spherical
surfaces. Radial modes are highly stable. Now if there is no means
of extracting energy from
the disk surface or the disk extends to the poles, then there will be
no net transport and the disk may be forced toward a state in which
radial convection can develop. This tendency may be present in
numerical simulations by SPB99, some of which show large mass
circulations with very small net mass flux.

Convection along gyrentropes is explicity suppressed, for example,
in the CDAF model by \citet{qua00b}, which extends
all the way to the polar axis.  We believe that this model is
physically implausible for two reasons. The first is that flows
that do not contain a central funnel have a singular velocity
along the axis $v_\phi(0)=1$. The rates of shearing, dissipation
and entropy production near the axis
will therefore diverge, driving a powerful outflow which
creates a funnel (\cf BB99). Because of the symmetry at this point, the viscous
stress at the pole cannot be balanced by convective stresses. The
second reason is that, unlike with the case  of accretion onto a star,
there is no means of supporting a column
of gas along the rotation axis above a black hole: the centrifugal
force acts along a perpendicular direction to the gravity and the pressure
force must vanish close the event horizon of the hole
\citep[\cf][, BB99]{nar97}.  Once the funnel exists, we assert that
convection will preferentially operate along gyrentropes, leading
to mass and energy loss from the funnel walls.

CDAFs also face secular difficulties related to the global mass
supply and energy flow.  In CDAFs, as in ADIOS models, the
accretion rate reaching the black hole is far smaller than the
mass supply at large radii.  However, unlike ADIOS models, there is
no means of escape for the vast majority of the supplied matter,
which does not make it to the black hole.   In the hydrodynamical
simulations by SPB99 and others, this is not an issue because these
calculations track the evolution of a finite torus; matter is not
continuously supplied.  (Moreover, the runs are not long enough to
track the evolution of the outer flow.) But in a CDAF with a continuous
mass supply, there is no alternative but to establish an increasing
reservoir of matter that is unable to accrete. Thus, CDAFs cannot
represent steady-state flows. Moreover, there is a continuous flow
of energy into the outer parts of a CDAF, where it will also build up
unless there is some escape route. It has been proposed that the outer
parts of ADAFs may adjust to radiate away this energy \citep{bal01,abr01},
but it has not been
demonstrated that this will happen naturally.  Another possibility is
that the energy flux will power a wind at some outer radius.  We contend
that this too is a less likely resolution than a scheme in which the
wind is released continously from all radii.

Even if CDAFs did not face these severe physical problems,
their existence seems unlikely to carry over to the MHD case,
which is more appropriate to astrophysical disks than the
hydrodynamical limit. The dominant instabilities in magnetized
disks are magnetorotational, rather than convective, and the resulting
turbulence is likely to transport angular momentum outward rather than
inward \citep{bal02}; but see \citet{nar02,igu02}
for an alternative view. If this is
the case, then the CDAF approach cannot be generalized to MHD flows.
However, we will argue in Paper II that the ADIOS approach does carry over.
Although MHD disks cannot attain marginal stability, they
nevertheless appear to develop a well-defined internal structure
(barytropicity, with rotation on cylinders), that provides an equivalent
autonomous constraint to our hydrodynamical assumption of
gyrentropicity. Thus, MHD disks admit well-defined circulation patterns
and outflows, which we will derive explicitly in Paper II.

\subsection{Weakly-Bound Disks}
\label{ssec:weak}

Finally, we comment on a class of non-radiative accretion models most
recently studied by \citet{pac98}, but which enjoyed wide
popularity in the early 1980's
\citep[\eg][and references therein]{jar80,pac80}.
In these models the angular momentum distribution
is prescribed, usually as a function of cylindrical radius along the
surface of the flow, between some inner radius $r_{\rm in}$ and some
outer radius $r_{\rm out}$. At $r\geq r_{\rm out}$ the flow is assumed
to match a thin Keplerian accretion disk, while at
$r \rightarrow r_{\rm in}$ the flow approaches a zero-pressure cusp
through which material accretes onto the black hole. For the interesting
limit of $r_{\rm out} \gg r_{\rm in}$ the gas in the advection-dominated
region is very weakly bound and one can show \citep{pac98}
that the cusp must lie close to the marginally bound orbit at $r = 4
m$ (for a Schwarzschild
black hole).  There is considerable latitude in the choice of angular
momentum distribution,
subject to the constraints that the angular momentum must be a
monotonically increasing function of $r$, and that the angular momentum
and binding energy must match the Keplerian values at both
$r_{\rm in}$ and $r_{\rm out}$.

One of the rationales for exploring these models
is that the generic form of the angular momentum distribution
is physically motivated, whereas viscous stress prescriptions are
highly uncertain. It is therefore worthwhile, Paczy\'nski argues, to
construct a model based on an assumed angular momentum law, and then see
what kind of viscous stress prescription is required to make it
self-consistent.  However, in his published work he does not discuss
the nature of these stress prescriptions.  We now show that
weakly bound disk models require very specific and (we believe)
implausible prescriptions for the viscous stress.

We restrict our attention to the limit $r_{\rm out} \gg r_{\rm in}$
and use the notation in \citet{pac98}. \citet{pac98} eq.~(25)
implies that the binding
energy is a monotonically increasing function of $r_s$, the
cylindrical radius measured along the surface. This implies the inequality
\begin{equation}
\label{P981}
\int^{r_s}_{r_{\rm in}} {1\over r^2} {d j_s^2 \over d r} dr
   < {2 j_{\rm out}^2\over r_{\rm out}^2 } \approx {1 \over 2
r_{\rm out} r_{\rm in} } j_{\rm in}^2
\end{equation}
for all $r_{\rm in} < r_s < r_{\rm out}$.  Integrating by parts,
and using the inequality $j_s \geq j_{\rm in}$, we finally obtain
\begin{equation}
\label{P982}
   j_s^2  < j_{\rm in}^2 \left( 1 + { r_s^2 \over 2 r_{\rm out}
r_{\rm in} } \right) .
\end{equation}
Thus, the angular momentum distribution must be a very weak
function of radius, $j \approx j_{\rm in}$ for $r_s$ smaller than the
geometric mean between $r_{\rm in}$ and $r_{\rm out}$.  In other words,
any acceptable angular momentum distribution for this kind of model
must be nearly constant in the inner part of the disk, while
increasing more steeply than Keplerian in the outer parts
(in order to satisfy the outer boundary condition).

Inequality (\ref{P982}) places severe constraints on the
viscous stress. Using the $\alpha$-model viscous couple assumed by
\citet{afs02}
which differs from our assumed torque (eq.~[\ref{magtorq}])
by a factor $\Omega / \Omega_{\rm Kep}$, and assuming an internal
sound speed $v_s \sim v_{\rm Kep}$ (appropriate for a thick disk),
we obtain an inflow speed $v_r$ given by
\begin{equation}
\label{P983}
{-v_r \over v_{\rm Kep}} \approx \alpha {j_s \over j_s -
j_{\rm in}} \left( - {d \ln\Omega\over d\ln r}\right) .
\end{equation}
Substituting from inequality (\ref{P982}) in the limit
$r_s \ll (r_{\rm in} r_{\rm out})^{1/2}$, we can write eq.~(\ref{P983})
in the form
\begin{equation}
\label{P984}
\alpha < {r_s^2 \over 8 r_{\rm in} r_{\rm out} }{-v_r \over v_{\rm Kep}} .
\end{equation}
The hydrostatic condition implies $- v_r/ v_{\rm Kep} \ll 1$,
placing a tight upper limit on $\alpha$. The limit is even more severe
if we use our expression (\ref{magtorq}) for the stress.

If $\alpha$ exceeds the limit given in eq.~(\ref{P984}), over any
range of radii, then the inflow will proceed on a free-fall timescale,
violating a principal assumption of the model.  This will presumably
continue until the angular momentum distribution relaxes to state
closer to Keplerian, in which case it will resemble one of the models
discussed above (i.e., an ADAF, CDAF, or ADIOS).  It will also evolve
toward a larger binding energy. We would, of course, argue that the only
self-consistent final state under these circumstances would be an ADIOS.

Physically, we expect the value of $\alpha$, or its equivalent in a
more realistic stress model, to be set by the microphysics of the
accretion process and not by the global boundary conditions ---
the latter presumably reflect the cooling function of the gas and/or
the nature of the gas supply at $r_{\rm out}$.  Numerical models of
MHD flows suggest that $\alpha$ may fall in the range $\sim 0.01-0.1$,
which is inconsistent with the limit derived above unless
$r_{\rm out} \la 100 r_{\rm in}$.  Observations of adiabatic accretion
flows suggest that $r_{\rm out}$ may be larger than $\sim 10^4 - 10^5$
times $r_{\rm in}$.  Such systems would have to have an effective
$\alpha < 10^{-5}$ in order to be described self-consistently
by a Paczy\'nski model. Thus, we conclude that weakly bound disk
models are unlikely to apply in many, if not most, cases of
astrophysical interest.

\section{Discussion}
\label{sec:dis}
We have attempted to flesh out the ADIOS concept by constructing
explicit,
two-dimensional models of adiabatic, accreting fluid.  Such flows are
strongly convective and should naturally evolve toward a state of
marginal stability.  We argue that the convective transport of
energy and angular momentum does {\it not} occur primarily
in the radial direction but, instead, proceeds primarily along surfaces that
connect the equatorial region to the disk ``surface" at high latitudes.
We have demonstrated how to construct models of two-dimensional disks
embodying these principles, and have then elaborated upon them
so as to include poloidal circulation, inflow and the formation of
fluid outflows.

Even in the context of purely fluid disks, our approximate
description of the flow can be challenged.  For example,
if the viscous stress, as measured by the parameter
$\alpha$, is not small, the perturbative ordering:
circular  speed $\gg$ convective speed $\gg$
meridional circulation/inflow speed will not be well-satisfied as we
require. Our thermal front model for producing the outflow is, likewise, an
oversimplification. It posits a site where entropy is produced
and simple conservation laws make explicit the connection between
the energy flow along the gyrentropes and the outflow. It seems reasonable
that the dissipation should be a strongly increasing function of the Mach
number and that it be efficient when the flow becomes
roughly sonic but this is not required. If the surface
dissipation is small, we suppose that the energy will
be advected by the circulation to large radius. Indeed,
as discussed in \S~\ref{ssec:numsim}, several simulations do seem to show
this behaviour which is not consistent with a self-similar
flow. Higher resolution
simulations that are run for long enough to achieve a stationary
flow will be needed to understand the flow of mass, angular momentum
and energy through the disk.

Another concern is that our calculation is expressly two-dimensional
and it is well-known that three-dimensional convection
in non-rotating fluid is quite different from its two-dimensional
counterpart. However, if our convection model is realistic, it automatically
leads to considerable smoothing over azimuth which ought to validate
the two-dimensional description. A related concern involves the unstable,
non-axisymmetric, global instabilities that can develop in the relativistic 
inner disk described in \S~\ref{ssec:reldisks} \citep{pap84}. As these modes 
rely upon reflection from the inner surface of the disk, they are unlikely 
to grow to a large amplitude in the presence of inflow \citep{beg84,bla87}.  

The most fundamental limitation concerns the explicit neglect of
magnetic field in our model. Beyond all reasonable doubt,
the torque in astrophysical accretion disks is hydromagnetic in origin
\citep{bal98}. If the magnetic fields remained small in strength and
length scale and
were locally dissipative, then our treatment could still have validity.
However, the evidence from numerical simulations is that none of these
conditions is well-satisfied in practice. In particular,
the relationship between the momentum transfer and the dissipation is
problematic.
With a simple, Newtonian viscosity, the connection is clear. There is a local
dissipation at a rate $\vec G\cdot\nabla\Omega$. Entropy is created
where the torque is applied. The introduction of convection into fluid models
complicates this linkage, because it involves the net bodily transfer of mass
carrying its own angular momentum and energy and transporting an additional
energy flux. This effectively makes the dissipation nonlocal.
As we shall discuss in Paper II, magnetized disks may provide
even more extreme examples of non-local dissipation. Unless
a magnetic turbulence spectrum is established and the energy
cascades down to a small
inner scale where it can be taken up by plasma, much of the
dissipation may be non-local. Indeed, there is plenty of
evidence that much of the dissipation in observed disks occurs
in a hot corona. (In considering this problem it is important to make the
distinction between ion-supported and radiation-supported disks. The former
require that local dissipation not heat the electrons to temperatures
where they can radiate efficiently, implying
either that most of the energy either goes into heating the ions
or that the energy is transported away.
By contrast, radiation-dominated accretion disks are subject
to the magnetorotational instability but the magnetic stress may
saturate at a lower level, perhaps validating a treatment
closer to our fluid models.)

The final missing ingredient is an allowance for
global time-dependence in the flow.
There are strong indications that observed
disks accrete episodically. Whether or not this happens in practice,
on all radial scales, depends upon the details of the
transition region. In \S~\ref{ssec:trans},
we argue that there could be a smooth, stationary  transition from a
radiative thin disk to an ADIOS.
However, if the outflow is less efficient
than we have assumed in carrying off the
energy, circulation of the excess energy
to the transition region will push it outward to progressively
larger radii until there is catastrophic cooling so that the thick disk
quickly shrinks. The whole pattern can repeat in a limit cycle.
However, as the inflow and dynamical timescales are always likely
to increase with radius, these variations should only cause
a slow secular change in the flow pattern at small radii, where most
of the accretion energy is released. In the language of our models this would
be manifested in a change in the entropy scale $s_0$  (\cf
\S~\ref{ssec:conserv}). The other way in which time-dependence could
be important is if there are local instabilities in the flow pattern.
In the case of adiabatic accretion disks the distinction between
viscous and thermal timescales is blurred. This, coupled with the
physical thickness of the disk, makes a formal instability calculation
more complicated and we advocate direct numerical
simulation as the means to explore stability of these flows. Few simulations
have been evolved long enough to draw strong conclusions about time-dependence.

We believe that the fluid dynamical analysis is valuable because it suggests a
``modular'' procedure for analysing numerical simulations of magnetized flows:
\begin{enumerate}
\item Determine the rule
that replaces gyrentropicity for determining the time-averaged disk
structure. We will argue in Paper II, on the basis of numerical
simulations and analytic calculations, that this principle
may be barytropicity, implying rotation on cylinders.
\item Separate the motion into relatively small scale interchanges -- with
scales no larger than the local pressure scale height, and larger
circulatory flows. The former can be analysed by computing
correlation functions involving velocity, pressure, density, etc.,
up to third order so as to understand the  time-averaged transport of
mass, angular momentum and energy as a function of position.
The latter will describe the global flow pattern.
\item Understand the surface boundary condition, specifically the
strength and character
of the MHD wind from the disk surface. Central to answering this question
is to decide whether or not large scale magnetic fields are generated,
as is observed in the high latitude wind from the quiet sun, or whether
the fields are tangled on small scale and behave as an anisotropic
gas.  This problem will be very hard to tackle as it combines both
local and global features.
\end{enumerate}
We shall address these issues further in Paper II.

Despite all of these shortcomings and concerns, we hope that the
models developed
in this paper will provide a guide to interpreting numerical simulations
and, ultimately, the increasingly detailed observations of radiation
from adiabatically
accreting black holes and neutron stars.

\section*{Acknowledgements}
We are indebted to many colleagues, notably Omer Blaes, Martin Rees
and Jim Stone,
for advice and encouragement. This work was supported in part by
NSF grants AST-9529170, AST-9529175 and AST-9876887, and NASA grants
NAG 5-2837 and NAG5-7007.
Much of the research reported here was carried out during 1999-2000
at the Institute of Astronomy at the University of Cambridge;
the Kavli Institute for Theoretical  Physics at the University of
California, Santa Barbara;
and the Institute for Advanced Study.  We thank the members of these
institutes for their hospitality.

\bsp

\end{document}